\documentclass[aip,jcp, twocolumns, reprint]{revtex4-1}
\usepackage[utf8]{inputenc}
\usepackage{amsmath}
\usepackage{amsfonts}
\usepackage{amssymb}
\usepackage{mathtools}
\usepackage{braket}
\usepackage{graphicx}
\usepackage{grffile}
\usepackage{siunitx}
\usepackage{rotating}
\usepackage{lipsum}  
\usepackage{array,multirow}

\begin{document}
	
	\title{The Adaptive Shift Method in Full Configuration Interaction Quantum Monte Carlo: Development and Applications} 
	
	\author{Khaldoon Ghanem}
	\affiliation{%
		Max Planck Institute for Solid State Research, Heisenbergstr. 1, 70569 Stuttgart, Germany
	}
	\email{k.ghanem@fkf.mpg.de}
	
	\author{Kai Guther}
	\affiliation{%
		Max Planck Institute for Solid State Research, Heisenbergstr. 1, 70569 Stuttgart, Germany
	}
	
	\author{Ali Alavi}
	\affiliation{%
		Max Planck Institute for Solid State Research, Heisenbergstr. 1, 70569 Stuttgart, Germany
	}
	\affiliation{%
		Dept of Chemistry, University of Cambridge, Lensfield Road, Cambridge CB2 1EW, United Kingdom
	}%
	
	\date{\today}

	\begin{abstract}
	In a recent paper, we proposed the adaptive shift method for correcting the undersampling bias of the initiator-FCIQMC. 
	The method allows faster convergence with the number of walkers to the FCI limit than the normal initiator method, particularly for large systems. In its application to strongly correlated molecules, however, the method is prone to overshooting the FCI energy at intermediate walker numbers, with convergence to the FCI limit from below.
	In this paper, we present a solution to the overshooting problem in strongly correlated molecules, as well as further accelerating convergence to the FCI energy.
	This is achieved by offsetting the reference energy to a value typically below the Hartree-Fock energy but above the exact energy. This offsetting procedure does not change the exactness property of the algorithm, namely convergence to the exact FCI solution in the large-walker limit, but at its optimal value greatly accelerates convergence. There is no overhead cost associated with this offsetting procedure, and is therefore a pure and substantial computational gain.  
	We illustrate the behavior of this {\em offset adaptive shift} method by applying it to the N$_2$ molecule, the ozone molecule at three different geometries (equilibrium open minimum, a hypothetical ring minimum, and a transition state) in three basis sets (cc-pV$X$Z, $X$=D,T,Q), and the 
	chromium dimer in cc-pVDZ basis set, correlating 28 electrons in 76 orbitals. 
	We show that in most cases the offset adaptive shift method converges much faster than both the normal initiator method and the original adaptive shift method.
	\end{abstract}
	\maketitle
	
	\section{Introduction}   
	The Schr\"odinger equation contains, in principle, all the information necessary for predicting the electronic structure  of quantum chemistry and solid-state physics.
	Yet solving the equation accurately remains an extremely challenging task except for small systems.
	The problem lies in the explosion of the size of the Hilbert space in which the solution lives.
	This size scales combinatorially with the number of electrons and the size of the basis set.
	As a result, an exact diagonalization method like Full Configuration Interaction (FCI) rapidly becomes prohibitive in the memory required for the entire wavefunction vector. 
	Full configuration interaction quantum Monte Carlo (FCIQMC) is a stochastic method that circumvents this problem by utilizing a sparse representation of the wavefunction using a set of signed walkers~\cite{Booth2009}.
	The walker dynamics is governed by the imaginary-time Schr\"odinger equation, and with sufficient walkers, converges in the long-time limit to the ground state of the system of interest.
	In its initiator variant, it has a comparatively low memory requirement, and the method is also highly-parallelizable and has been successfully applied not only to molecules but also a wide variety of systems including solids~\cite{Booth2013}, the uniform electron gas~\cite{Shepherd2012, Luo2018} , and the Hubbard model~\cite{Dobrautz2019}.
	The utility of the method is not restricted to ground states but has been extended in several different ways, including to excited states\cite{Blunt2015b}, calculating properties by sampling reduced density matrices\cite{Overy2014}, obtaining electronic spectra\cite{Guther2018} and full spin-adaptation, targeting specific spin states, using the graphical unitary group approach~\cite{Dobrautz2019b}. 

	Although the original FCIQMC algorithm can converge to the exact ground state, its convergence requires a minimum number of walkers in order to maintain stable dynamics and in order for the sign structure of the wavefunction to emerge~\cite{Spencer2012}.
	This minimum number depends on the system under study and is a fraction of the full size of the Hilbert space but still scales with it, making the method impractical for large systems.
	The initiator approximation (i-FCIQMC) solves this issue at the cost of introducing a systematically-improvable bias.
	In the initiator approximation, the dynamics of the walkers is constrained such that only walkers on determinants with enough population (called initiators) are allowed to spawn new walkers onto empty determinants.
	Spawns of a non-initiator to the empty determinants are rejected.
	This constrained dynamics prevents non-initiators, which are potentially sign-incoherent, from propagating noise and contaminating the wavefunction.
	Consequently, the simulation remains stable at any number of walkers, allowing i-FCIQMC to tackle much larger systems than the original FCIQMC algorithm.
	The main drawback is that the initiator criterion effectively scales down certain off-diagonal elements of the Hamiltonian, leading to an undersampling bias of the non-initiators.
	
	In a recent paper~\cite{Ghanem2019}, we proposed a remedy to this bias by giving each non-initiator determinant its own local shift, as a proportionately-scaled value of the overall global shift.
	The scale factors are based on the weighted acceptance probability of the non-initiator successfully spawning a walker, where the weights are computing using perturbation theory.
	We showed that this method reduces the initiator bias significantly, and leads to convergence to near-FCI energies at a much lower number of walkers than the normal initiator method.
	However, it was also found that, in some cases, the method rather overcorrects the initiator bias, leading to an overshoot in the energy estimates and a convergence from below.
	As a result, a global way of controlling the strength of the correction is desired.
	In this paper, we present a simple way of tuning the adaptive shift correction that results in a faster converge of the method than either the normal initiator and the original adaptive shift.
	
    The paper is structured as follows. We first review the FCIQMC in both its original formulation and its initiator approximation and discuss the bias it introduces.
	We then explain the adaptive shift method and how it rectifies this bias.
	We introduce an offset version of the adaptive shift as a simple method which gives further control on the extent to which the local shift at each non-initiator
    is modified, and demonstrate its effect using the example of N\textsubscript{2} molecule, at equilibrium and stretched geometries in the cc-pVDZ and cc-pVQZ basis sets.
	Next, we show results using this extended version of the adaptive shift for two challenging highly multi-configurational systems: ozone, and the chromium dimer.	We conclude with a summary and a discussion.
	
	\section{FCIQMC and The Initiator Approximation} 
	FCIQMC is a stochastic projective method in the space of Slater determinants~\cite{Booth2009}.
	As a projective method, it starts from some initial state with non-zero overlap with the ground state and evolves it according to the imaginary-time Schr\"odinger equation
	\begin{equation}
		-\frac{d}{d\tau}\ket{\psi(\tau)} =  \left(\hat{H}-E\right) \ket{\psi(\tau)}
	\end{equation}
	Setting $E$ equal to the ground state energy $E_0$, contributions from excited states decay exponentially and the wavefunction  $\ket{\psi(\tau)}$ approaches the ground state in the long-time limit.
	The ground state energy is not known a priori, so it must be estimated from the wavefunction itself, as will be discussed later.
	
	As a determinant method, FCIQMC solves the Schr\"odinger equation in the Hilbert space spanned by determinants constructed from a finite orthonormal basis of single-particle orbitals.
	Let $C_i \coloneqq \braket{D_i|\psi}$ be the projection coefficient on determinant $\ket{D_i}$, known as Configuration Interaction (CI) coefficient, the Schr\"odinger equation in this space reads
	\begin{equation}
		\label{eq:ci_eq}
		-\frac{d}{d\tau}C_i(\tau) =  \left[H_{ii} - E\right] C_i(\tau) + \sum_{i\neq j} H_{ij} C_j(\tau)
	\end{equation}
	where $H_{ij} \coloneqq \braket{D_i|\hat{H}|D_j}$ is the Hamiltonian matrix in the CI basis.
	Deterministic integration of the above equation  would require storing the entire CI vector, which is prohibitive except for small systems; the number of determinants scales combinatorially with the number of electrons and the size of the single-particle basis.
	
	This memory bottleneck is reduced in FCIQMC by a stochastic sampling of the wavefunction such that only a fraction of the full Hilbert space is simultaneously occupied. 
	The wavefunction is encoded using a population of walkers, each of which has either a positive or negative sign and occupies one determinant only.
	Walkers can die or duplicate on their own determinants, and they can spawn walkers on other determinants connected by non-zero matrix elements.
	A CI coefficient of a determinant $C_i$ is represented by the signed sum of walkers on that determinant $N_i$. 
	If the population of walkers is then updated randomly such that the average change at each time step respects the Schr\"odinger equation  i.e.
	\begin{equation}\label{eq:master}
		-{\frac{d }{d \tau}} \overline{N_i}(\tau) = \left(H_{ii} - E\right) N_i(\tau) + \sum_{i\neq j} H_{ij} N_j(\tau) \;, 
	\end{equation}
	then due to linearity, the average values themselves satisfy the Schr\"odinger equation
	\begin{equation}
		-\frac{d }{d \tau} \overline{N_i}(\tau)= -\left(H_{ii} - E\right) \overline{N_i}(\tau) + \sum_{i\neq j} H_{ij} \overline{N_j}(\tau) \;,
	\end{equation}
	and we get CI coefficients as averages of the walker populations
	\begin{equation}
		C_i(\tau) = \overline{N_i}(\tau)\;.
	\end{equation}

	In order for the linearity argument to hold, the energy estimate $E$ must be independent of the walker population.
	In practice, however, a decomposition into a fixed Hartree-Fock energy term $E_{HF}$ and a varying shift term $S(\tau)$ is used
	\begin{equation}\label{eq:hf_shift}
		E(\tau) = E_{HF} + S(\tau)\;,
	\end{equation}
	where the shift is updated periodically to keep the total number of walkers roughly constant.
	As a result, there is a weak correlation between the shift and the walker population leading to what is known as  population bias~\cite{Vigor2015}.
	This bias is typically very small and negligible in comparison to the statistical error that inherently exits in estimates of FCIQMC.
	In the long-time limit, the shift $S$ fluctuates around the correlation energy and $E$ around the total energy.
	
	The master equation~\eqref{eq:master} is carried out via three steps of walker dynamics:
	a spawning step corresponding to the non-diagonal elements of the Hamiltonian, a death step corresponding to the diagonal elements, and an annihilation step, where walkers of opposite signs on the same  determinant are canceled and removed.
	The details of these steps are explained in Ref.~\cite{Booth2009}.
	Several technical improvements of the original algorithm have been proposed since then:
	floating-point weights for walkers, energy estimation using trial wavefunctions, semistochastic projection~\cite{Petruzielo2012, Blunt2015}, efficient parallelization~\cite{Booth2014}, and efficient excitation generation~\cite{Holmes2016, Neufeld2019}.
	These improvements help reduce the statistical errors and the computational cost significantly but do not fundamentally change the dynamics of the walkers.
	
	Maintaining stable dynamics in the original FCIQMC requires a system-dependent minimum number of walkers.
	Below this number, the signs of low-populated determinants fluctuate randomly, leading to sign-incoherent noise.
	This noise then propagates via spawning and dominates the wavefunction preventing it from converging to the ground state.
	Although the minimum number of walkers is typically less than the full size of the Hilbert space, it still scales exponentially with the system size.
	This is the manifestation of the sign-problem in FCIQMC~\cite{Spencer2012}.
	
	The initiator approximation (i-FCIQMC) overcomes the issue by constraining the dynamics of the walkers~\cite{Cleland2010}.
	In this method, walkers residing on determinants with enough population (denoted as $n_a$ and typically chosen to be three) are called initiators.
	These walkers are the only ones allowed to spawn onto empty determinants.
	The other walkers (non-initiators) can spawn only onto an already-occupied determinant. 
	In essence, i-FCIQMC is modifying the non-diagonal matrix elements of the master equation~\eqref{eq:master}  as following
	\begin{equation}\label{eq:init_master}
		-{\frac{d }{d \tau}} \overline{N_i}(\tau) = \left[H_{ii} - E_{HF} - S(\tau)\right] N_i(\tau) + \sum_{i\neq j} \tilde{H}_{ij}(\tau)  N_j(\tau) \;, 
	\end{equation}
	where
	\begin{equation}\label{eq:init_matel}
		\tilde{H}_{ij}(\tau) = \begin{cases}
			H_{ij} \quad &\text{if }  |N_j(\tau)| > n_a \text{ or } N_i(\tau) \neq 0\\
			0 \quad &\text{ otherwise}
		\end{cases}
	\end{equation}
	This prevents non-initiators, whose sign may not be stable, from propagating sign-incoherent noise and thus allows convergence with arbitrarily small number of walkers. 
	The price, however, is introducing an initiator bias that gets smaller as the number of walkers increases.
	In the limit of a large population, all determinants become occupied and the full method is recovered, yielding exact results.

	Nevertheless, it is of interest to be able to mitigate the effects of the initiator bias, either by providing estimates of the initiator bias at a given walker number, or else by accelerating the convergence with walker number towards the exact FCI limit. Regarding the first strategy, perturbative estimates of the initiator bias can be made, to second \cite{Blunt2018} or more recently third-order \cite{Tenno2020} Epstein-Nesbet perturbation theory, inspired by similar PT2 corrections to selected CI methodologies \cite{Sharma2017, Garniron2017}.  Preconditioning methods \cite{Blunt2019} help in reducing the inherent noise in the PT2 estimators, although the replica-trick \cite{Overy2014} is necessary to obtain unbiased energy estimates. Regarding the second strategy of accelerating convergence to the exact FCI limit, the adaptive shift (AS) methodology \cite{Ghanem2019} has been proposed, which aims at improving the sampled wavefunction, not just the energy. Such improved wavefunctions allow calculation of properties \cite{Thomas2_2015} via reduced density matrices, including electronic gradients in stochastic CASSCF procedures \cite{Thomas2015,LiManni2016}. It has been shown that such improved wavefunctions (for example at the level of triple and quadruple excitations) can be successfully used in other methods, such as externally corrected and tailored Coupled-Cluster theory \cite{Deustua2018, Vitale2020,Eriksen2020}.  The AS methodology is 
	closer in spirit to coupled electron-pair approximations (CEPA) \cite{Tenno2017} and also down-folding concepts \cite{Lowdin1951}, effectively modifying the diagonal matrix elements via local shifts at the determinants at which the initiator approximation is occurring.           
	
	\section{The Adaptive Shift Method} 
    The initiator bias is the result of the undersampling of non-initiator determinants. 
	From Eq.~\eqref{eq:init_matel}, we see that Hamiltonian matrix elements between any two initiators or between an initiator and a non-initiator stay unmodified. 
	However, matrix elements between any two non-initiators get effectively reduced in the long-run, because whenever one of them is zero, the other cannot spawn onto it. 
	
	To see how one might rectify such a bias, let us  first consider the following simpler truncation criterion:
	Spawns are allowed only to a pre-specified set of determinants $\mathcal{A}$, while spawns to other determinants  $\mathcal{R}$ are rejected.
	The rejection of spawns onto set $\mathcal{R}$ biases the walker population in the set $\mathcal{A}$ due to the lack of subsequent back-spawns from $\mathcal{R}$ to $\mathcal{A}$.
	Nevertheless, one can still, in principle, recover the exact dynamics of the set $\mathcal{A}$, if the exact wavefunction amplitudes $C^\star_i $ are known.
	This can be achieved by rewriting the master equation \ref{eq:master} such that each determinant in $\mathcal{A}$ gets its own \emph{local shift} $S_i^\star$
	\begin{equation}
	-{\frac{d }{d \tau}} \overline{N_i}(\tau) = \left[H_{ii} - E_{HF} - S^\star_i(\tau) \right] N_i(\tau) + \sum_{\substack{j\in \mathcal{A}  \\ j\neq i }} H_{ij} N_j(\tau)\;,
	\end{equation}
	where the back-spawns of the set $\mathcal{R}$  are now folded into the local shift 
	\begin{equation}
	S^\star_i(\tau) \coloneqq S(\tau) -\sum_{j\in \mathcal{R} } H_{ij} \frac{C_j^*}{C_i^*}\;.
	\end{equation}
	In practice, this formal expression of the local shift should be approximated by some heuristic estimate of the missing back-spawns.
	
	Motivated by the above argument, the adaptive shift method uses the idea of local shifts to compensate for the undersampling bias of the initiator criterion.
	The truncation in the initiator method is, however, more intricate and differs in two aspects from above.
	First, the set of rejected determinants $\mathcal{R}$ is dynamically changing from one time step to the next. As a result, the local shifts should also be updated dynamically to reflect this dynamic truncation.
	Second, only the spawns of non-initiators onto $\mathcal{R}$ are rejected, while initiators can spawn freely.
	Therefore, we need to keep the shifts of initiators unmodified.
	
	In Ref.~\cite{Ghanem2019}, we proposed using local shifts $S_i(\tau)$ that are fractions of the total shift $S(\tau)$
	\begin{equation}\label{eq:local_shift}
	S_i(\tau) = f_i  \times S(\tau)\;,
	\end{equation}
	where the fraction $f_i$ should be proportional to how severely determinant $\ket{D_i}$ is affected by the undersampling.
	A systematic and computationally-cheap way of assigning these factors is as follows.
	We associate a positive weight $w_{ij}$ with each attempted spawn from determinant $\ket{D_i}$ to determinant $\ket{D_j}$, and monitor which of these spawns are accepted and rejected due to the initiator criterion.
	The factor $f_i$ of determinant $\ket{D_i}$ is then computed as the ratio of accepted weights over all weights
	\begin{equation}\label{eq:as_fval}
	f_i=\frac{ \sum_{j\in accepted} w_{ij} }{ \sum_{j\in all} w_{ij} }\;.
	\end{equation}
	Regardless of the exact form of the weights, the above expression ensures that initiators get the total shift automatically.
	Besides, the local shift of a non-initiator monotonically approaches the total shift as its local Hilbert space becomes more populated. 
	This guarantees that we restore the original master equation of FCIQMC in the large walker limit.
	
	In this study, we use weights $w_{ij}$ derived from perturbation theory~\cite{Lowdin1951} where the first-order contribution of $\ket{D_i}$ to the amplitude of $\ket{D_j}$ is used as a weight for spawns from $\ket{D_i}$ to $\ket{D_j}$
	\begin{equation}
		w_{ij} = \frac{|H_{ij}|}{H_{jj}-E_0}\;.
	\end{equation}
	We found these weights to be particularly suitable for ab initio systems because they account for the nonuniform matrix element magnitudes $|H_{ij}|$ in these Hamiltonians and the resulting non-equal contributions in a local Hilbert space.
	
	\subsection{Fine Tuning The Adaptive Shift}
	In Ref.~\cite{Ghanem2019}, we showed that the adaptive shift method is effective in correcting the undersampling bias caused by the initiator criterion.
	As the number of walkers increases, energy estimates using the adaptive shift method typically converge much faster than the normal initiator method .
	Still, we found that for some systems, the energy converges from below, indicating that the adaptive shift is rather over-correcting.
	It is then desirable to be able to tune down this correction.
	
	A key observation is noting that the definition of the total shift $S$ is not unique. 
	In Eq.~\eqref{eq:hf_shift}, the shift is taken with respect to the Hartree-Fock energy, but we can equally take it with respect to any other reference energy $E_\text{Ref}\geq E_0$.
	Without the adaptive shift method, this change of reference does not affect the results, and the new shift would fluctuate around the residual energy $E_0-E_\text{Ref}$ instead of the correlation energy. 
	The choice of Hartree-Fock energy $E_\text{HF}$ as a reference is driven by convenience since its shift can be used as an independent estimator of the correlation energy.	
	Using the adaptive shift method, however, the choice of the reference energy affects the amount of death rate that gets scaled down by the factors $f_i$.
	In the extreme case when $E_\text{Ref} = E_0$, the total shift $S(\tau)$ fluctuates around zero and the modification factors have little to no effect on the result.
	
	To account for this degree of freedom, we introduce an offset parameter $\Delta\coloneqq E_\text{Ref} -E_\text{HF} \geq E_\text{correlation}$ and redefine the local shift as following
	\begin{equation}\label{eq:local_shift_offset}
		S_i(\tau) = \Delta + f_i  \times \left[S(\tau)-\Delta\right]\;.
	\end{equation}
	This new parameter allows us to control the overall strength of the adaptive shift correction. 
	When $\Delta=0$, this definition coincides with the original definition of the local shift \eqref{eq:local_shift},  and the correction is applied fully. 
	When $\Delta=E_\text{correlation}$,  the adaptive shift has practically no effect on the result, and we are back to the normal initiator. We therefore expect appropriate values of the offset $\Delta$ to be in the range $E_\text{correlation}<\Delta<0$, and with a guess of the correlation energy (obtained for example from MP2, or coupled cluster or other sources), an educated guess for a good value of $\Delta$ can easily be made. As will transpire from the calculations we will report, 
	$\Delta\approx E_\text{correlation}/2$ seems to be a universally good value. 
	Note that for initiators, i.e., when $f_i=1$, the local shift is independent of the offset and always equals the total shift.
	
	\subsection{Illustrative Example}
	\begin{figure}[h]
	\center
	\includegraphics[width=\columnwidth]{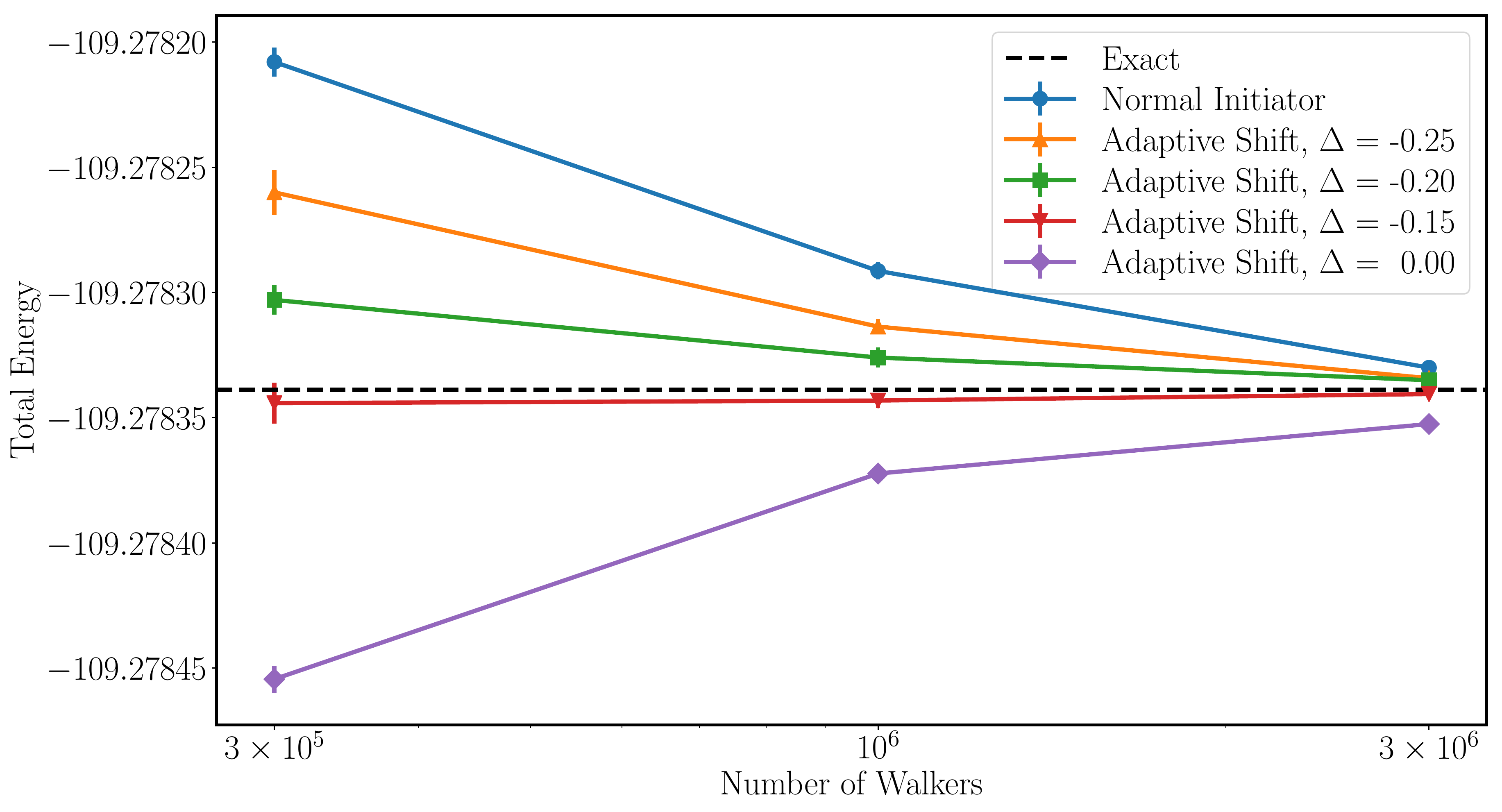}
	\caption{\label{fig:n2_2.118_vdz}
		N\textsubscript{2} in cc-pVDZ basis and near-equilibrium geometry: $\SI{2.118}{\bohr}$.
	}
	%\end{figure}
	%\begin{figure}
	\center
	\includegraphics[width=\columnwidth]{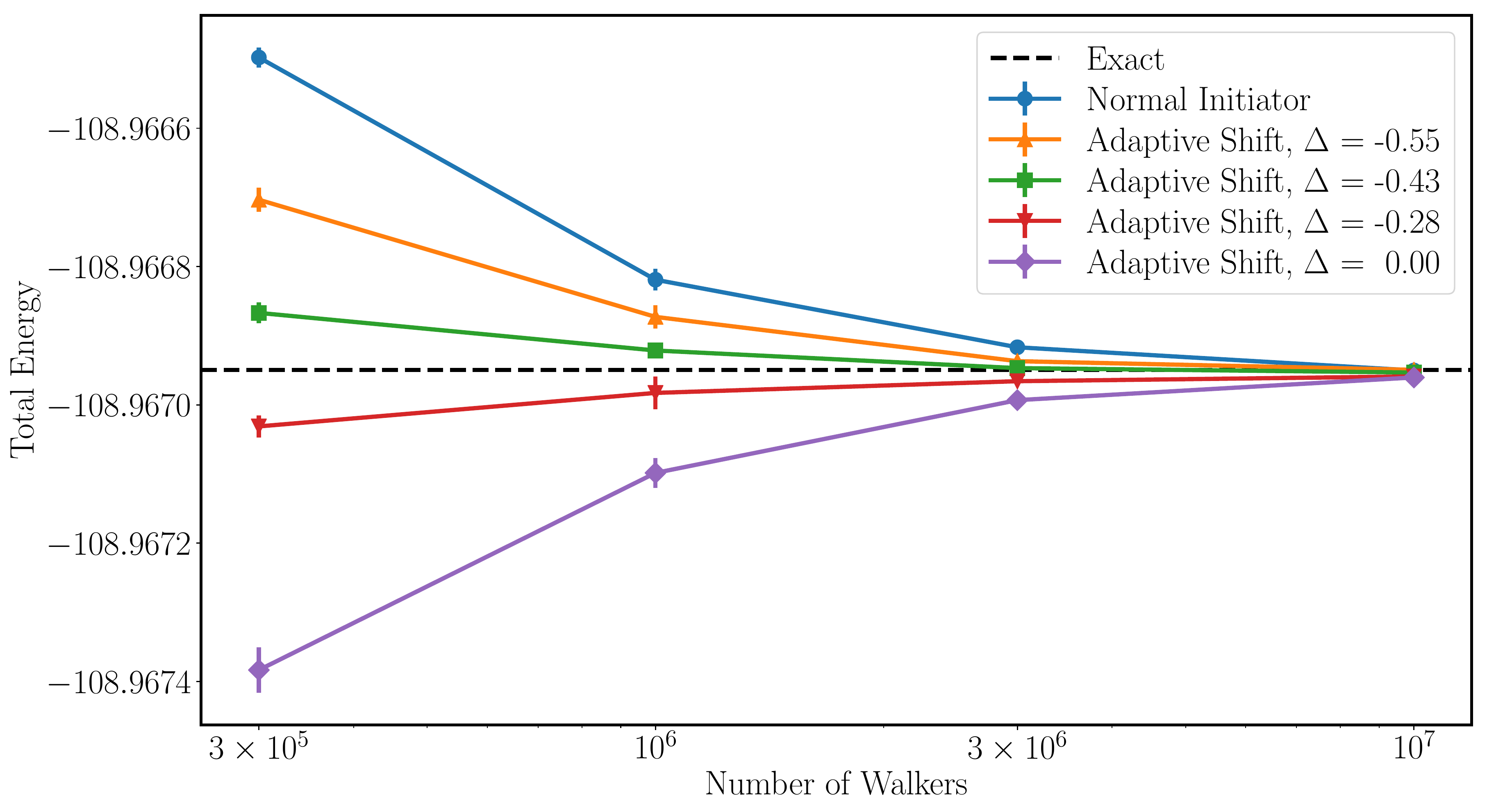}
	\caption{\label{fig:n2_4.2_vdz}
		N\textsubscript{2} in cc-pVDZ basis and stretched geometry: $\SI{4.2}{\bohr} $.
	}
	%\end{figure}
	%\begin{figure}
	\center
	\includegraphics[width=\columnwidth]{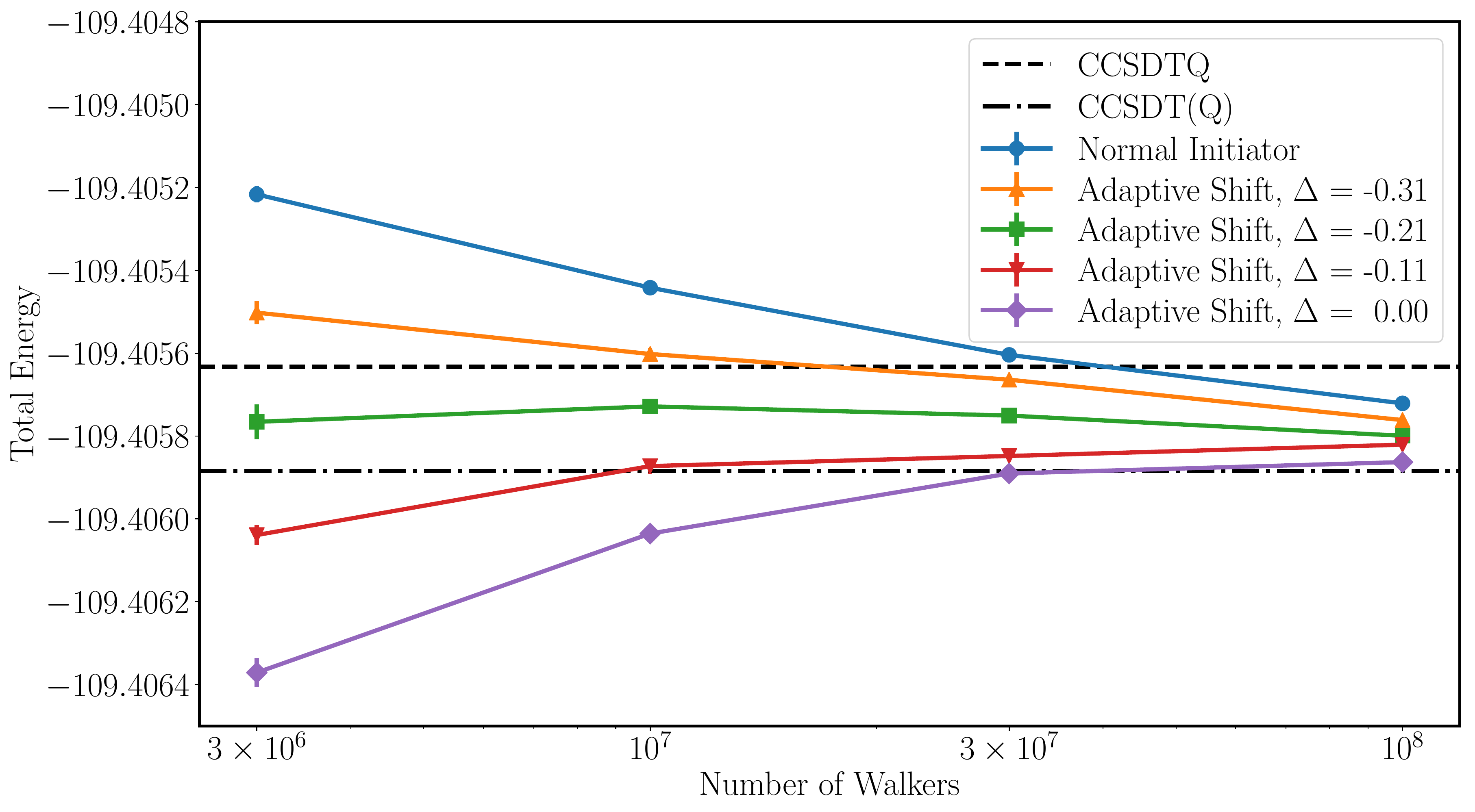}
	\caption{\label{fig:n2_eq_vqz}
		N\textsubscript{2} in cc-pVQZ basis and equilibrium geometry: $\SI{1.0977 }{\angstrom}$.
	}
\end{figure}	
	To illustrate the effect of varying the offset parameter of the adaptive shift, we consider N\textsubscript{2} molecule in cc-pVDZ basis for a near-equilibrium geometry ($\SI{2.118}{\bohr}$) and a stretched geometry ($\SI{4.2}{\bohr} $). This system is a toy problem  (10 electrons in 26 orbitals) where the exact result is known, and the normal initiator method already converges to chemical accuracy using less than $10^5$ walkers.

	For the near-equilibrium geometry, the initiator method using $n_a=3$ and $3\times 10^5$ walkers gives $-109.278208(6)$ a.u., which is $0.131$  mH above the exact energy of $-109.278339$. 
	Applying the full adaptive shift correction (i.e., with offset $\Delta=0$), we get $ -109.278454(5)$, which is below the exact energy by a similar amount ($0.115$ mH). 
	As the number of walkers is increased to $3$M, the initiator method converges from above, while the full adaptive shift converges from below,  bracketing the exact energy to within $0.02$ mH.
	Clearly, the full adaptive shift is over-correcting here, so we next apply the adaptive shift with an offset  $\Delta=-0.15$, which is around half the correlation energy.
	Using this offset, we get $-109.278344(8)$ using only $3\times 10^5$ walkers,  matching the exact energy within $0.005$  mH.
	As the number of walkers increases, the energy stays constant within the statistical error, which shows that the adaptive shift correction for this offset cancels almost exactly the initiator bias.
	Lowering the offset further to $\Delta=-0.2$ and $\Delta=-0.25$ leads to less correction and thus higher energies.
	The results are shown in Fig.~\ref{fig:n2_2.118_vdz}.
	
	We observe similar behavior for the stretched geometry as we vary the offsets and the number of walkers (See Fig.~\ref{fig:n2_4.2_vdz}).
	The initiator method converges from above, while the full adaptive shift ($\Delta=0$) overshoots and converges from below.
	The errors of both methods are larger here than for the previous geometry.
	For example, using $3\times 10^5$ walkers, the initiator method is $0.449$ mH above the exact energy of $-108.966950$, and the full adaptive shift is $0.433$ mH below it.
	Setting the offset to $\Delta=-0.28$ reduces the overshoot of the adaptive shift significantly, yet the results still converge from below.
	By lowering the offset to $\Delta=-0.43$, the results now converge from above.
	Using these two offsets, the exact energy  is bracketed to within $0.16$ mH at already $3\times 10^5$ walkers. 
	For the largest number of walkers of $10$M, all results are within $0.01$ mH of the exact energy. Again for this stretched system, an offset
	of roughly half of the correlation energy (-0.744 mH) seems to be an appropriate value to select.  
		
	We also considered the equilibrium geometry ($\SI{1.0977 }{\angstrom}$) in the larger cc-pVQZ basis (108 orbitals) whose results are shown in Fig.~\ref{fig:n2_eq_vqz}.
	Using $3$M walkers, the normal initiator gives $-109.40522(2)$ a.u., which is $0.41$ mH above the CCSDTQ result of $-109.405633$. 
	On the other hand, the full adaptive shift gives $-109.40637(4)$ a.u. which is  $0.49$ mH below CCSDT(Q) result of   $-109.405885$.
	Increasing the number of walkers to $100$M, the normal initiator energy comes down below the CCSDTQ result to $-109.405721(6)$, while the full adaptive shift energy goes above the CCSDT(Q)  to  $-109.405863(9)$.
	We applied the offset adaptive shift method with three different offsets.
	Using $\Delta=-0.11$, the results converge from below faster than the full adaptive shift to the value $-109.405821(6)$ a.u.
	Using $\Delta=-0.31$, the results converge from above faster than the normal initiator to the value $-109.405761(7)$ a.u.
	The best correction is obtained using $\Delta=-0.21$, where the energy stays roughly constant in the middle as the number of walkers increases.
	With $3$M walkers, this offset gives energy $-109.40577(4)$ a.u., which is only $0.02$ mH away from the value $-109.405799(7)$ a.u. obtained with the much larger population of $100$M walkers. Since the total correlation energy is approximately $-0.415$, this again implies that a good value 
	of the offset is simply half of the total correlation energy.

	\section{Results}
	In the following, we report the results of the adaptive shift for two systems well-known to be challenging and have been the subject of recent high-level calculations: the ozone molecule and the chromium dimer. In the case of the ozone molecule, a recent CEEIS study by Ruedenberg et al \cite{Theis2016} concluded that accurate energy differences in this system require CI expansions to at least sextuple excitations. High-order coupled cluster calculations (Table I) indicate significant differences between the energy gaps predicted by CCSDT, CCSDT(Q), and CCSDTQ, this being another indication of the intricate level of correlation necessary to describe this system. 
	%and it is difficult to go to larger basis sets.  

    The calculations were done using the {\tt NECI} code, which provides a state-of-the-art and efficient implementation of FCIQMC~\cite{Guther2020}.
	In all calculations, the initiator threshold is set to $n_a=3$. 
	The calculations start with 100 walkers on the Hartree-Fock (HF) determinant and grow the population with zero shift till the target number of walkers $N_w$ is reached. 
	The shift is then varied to maintain a stable number of walkers.
	When the adaptive shift method is applied, the local shifts are used only after the target population is reached, and the shift starts varying. 
	After $1000$ iterations of reaching the target population, a semistochastic space $\mathcal{D}$ and a trial wavefunction space $\mathcal{T}$  are constructed out of the topmost occupied determinants.
	Energy estimates are computed by projecting on the trial wavefunction.
	Tables of the results can be found in the supplementary material file.
	
	\subsection{Ozone}
		\begin{figure}
	\center
	\includegraphics[width=\columnwidth]{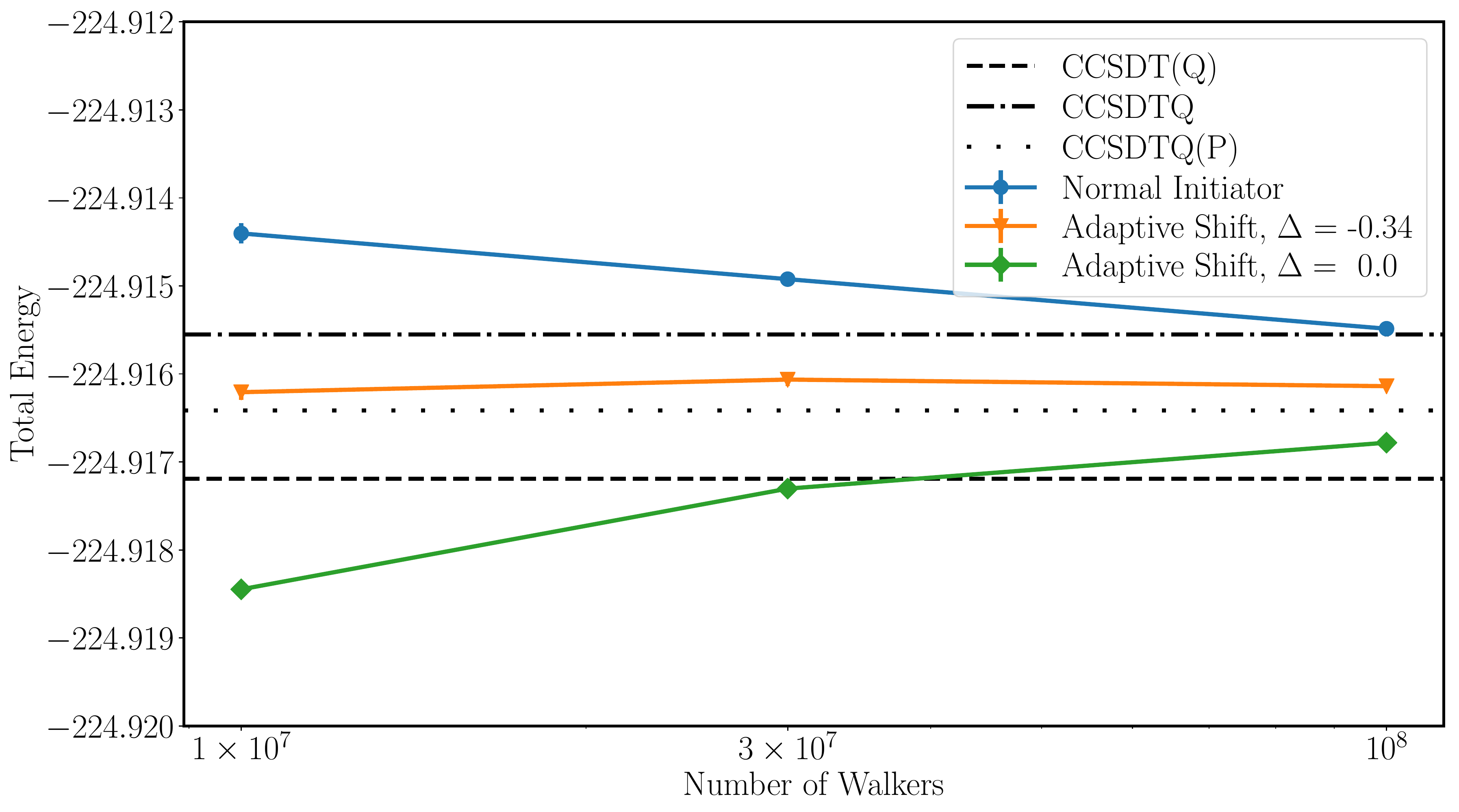}
	\caption{\label{fig:ozone_om_vdz}
		Ozone in cc-pVDZ basis and OM geometry.
	}
	%\end{figure}
	%\begin{figure}
	\center
	\includegraphics[width=\columnwidth]{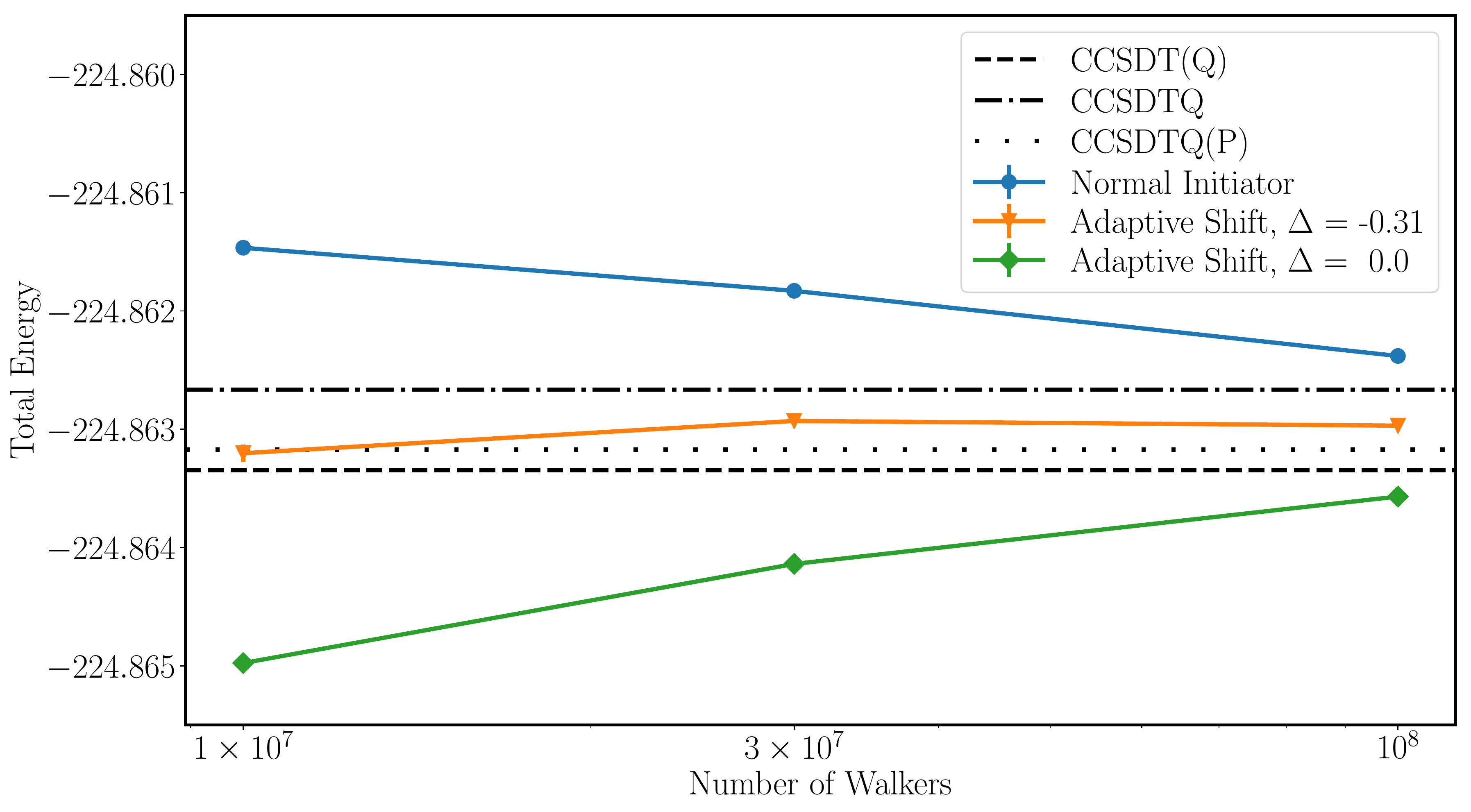}
	\caption{\label{fig:ozone_rm_vdz}
		Ozone results in cc-pVDZ basis and RM geometry.
	}
	%\end{figure}
	%\begin{figure}
	\center
	\includegraphics[width=\columnwidth]{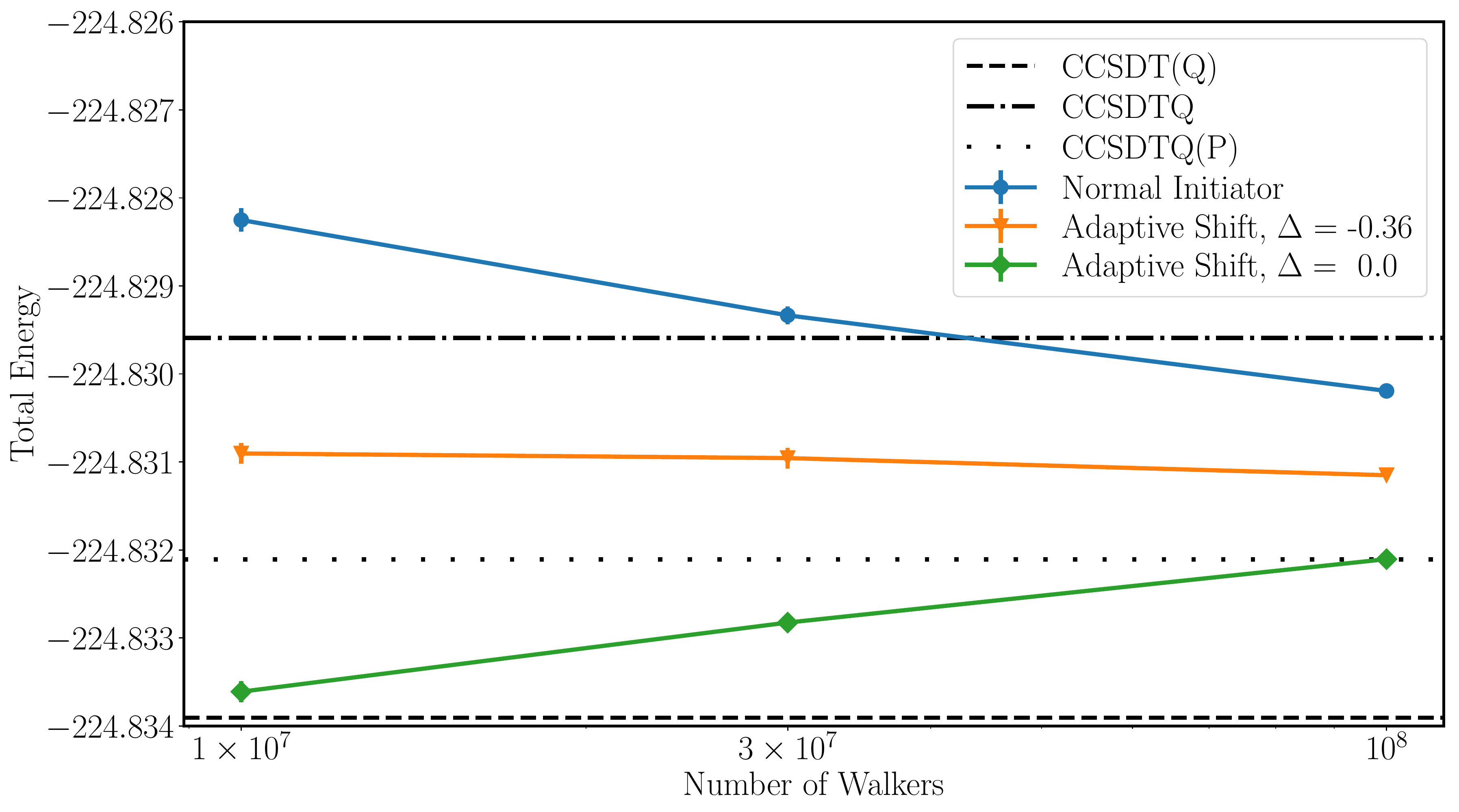}
	\caption{\label{fig:ozone_ts_vdz}
		Ozone  in cc-pVDZ basis and TS geometry.
	}
\end{figure}	
	The ozone molecule was treated in three basis sets: cc-pVDZ (39 orbitals), cc-pVTZ (87 orbitals) and cc-pVQZ (162 orbitals) and three different geometries: an open minimum  (OM), a meta-stable ring minimum (RM), and a transition state (TS)  between these two minima, correlating 18 valence electrons. RHF orbitals were used. 
	For all cases, we ran normal initiator calculations, full adaptive shift calculations, and offset adaptive shift calculations with offsets chosen to be roughly half of the correlation energies.
	We set the sizes of the semistocastic and trial spaces respectively as following: $|\mathcal{D}| =  1000$ and $|\mathcal{T}| = 100$.
	As the number of walkers increases, in all cases, the initiator calculations converge from above, while the full adaptive shift calculations converge from below. 
	The offset adaptive shift energy is bracketed from above by the normal initiator and below by the full adaptive shift method and gives the most well-converged results.

	For the cc-pVDZ basis, we ran calculations using $10$M, $30$M, and $100$M walkers.
	For the OM geometry, the offset adaptive shift energy is right between the CCSDTQ result of $-224.91555$ and the CCSDTQ(P) result of $-224.91641$ with the value $-224.91614(3)$  obtained using $100$M walkers (Fig.~\ref{fig:ozone_om_vdz}).
	For the RM geometry,  the offset adaptive shift energy using $10$M walkers coincides with the CCSDTQ(P) result of $-224.86317$ but then raises to $-224.86297(4)$  using $100$M walkers.
	This is about $0.3$ mH below CCSDTQ result of $-224.86266$  (Fig.~\ref{fig:ozone_rm_vdz}).
	For the TS geometry, the offset adaptive shift energy stays relatively stable as the number of walkers increases with value $-224.83115(4)$  obtained using $100$M walkers. This value lies again between the CCSDTQ result of $-224.82959$ and the CCSDTQ(P) result of $-224.83211$  (Fig.~\ref{fig:ozone_ts_vdz}).
	
\begin{figure}
		\center
		\includegraphics[width=\columnwidth]{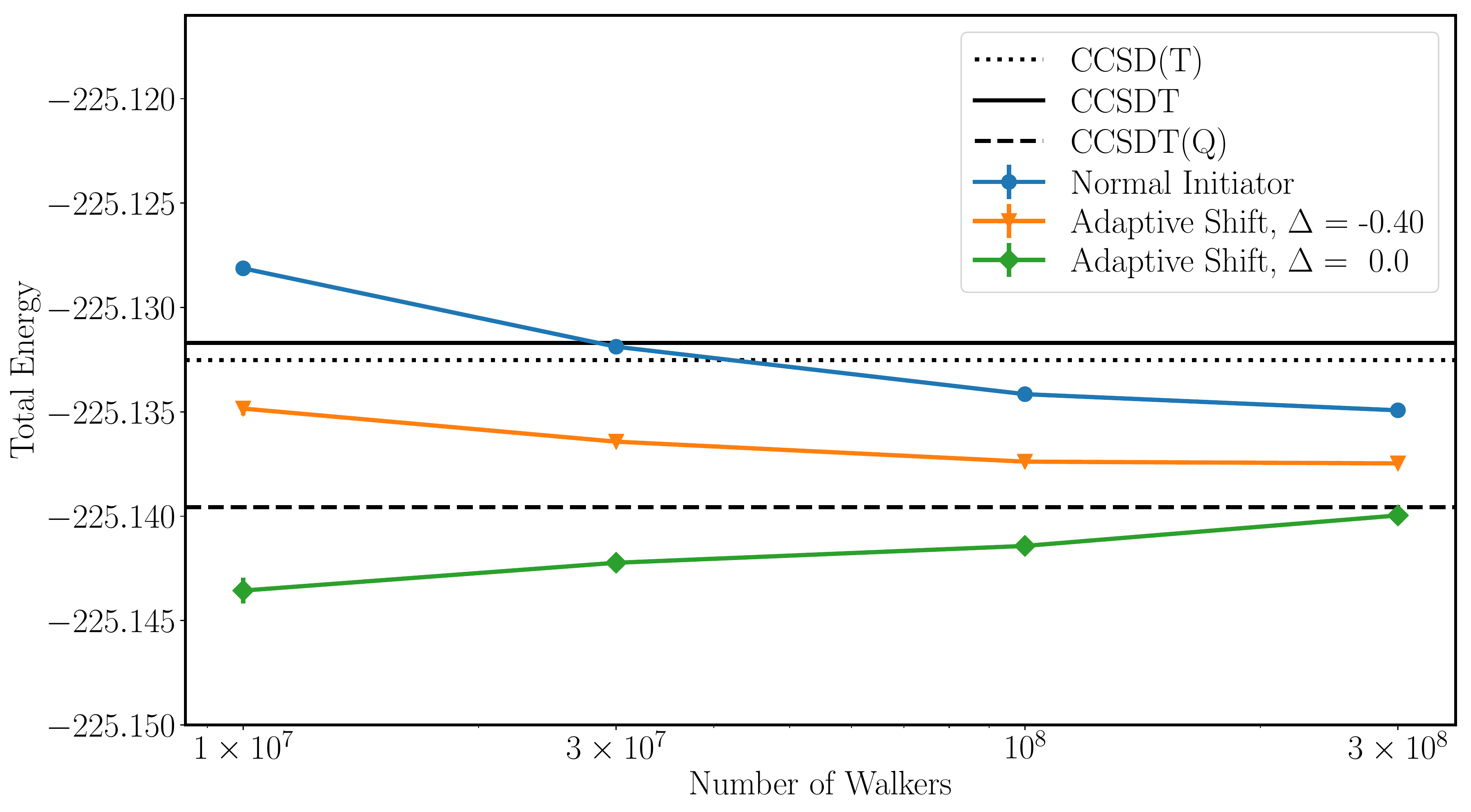}
		\caption{\label{fig:ozone_om_vtz}
			Ozone in cc-pVTZ basis and OM geometry.
		}
		%\end{figure}
		%\begin{figure}
		\center
		\includegraphics[width=\columnwidth]{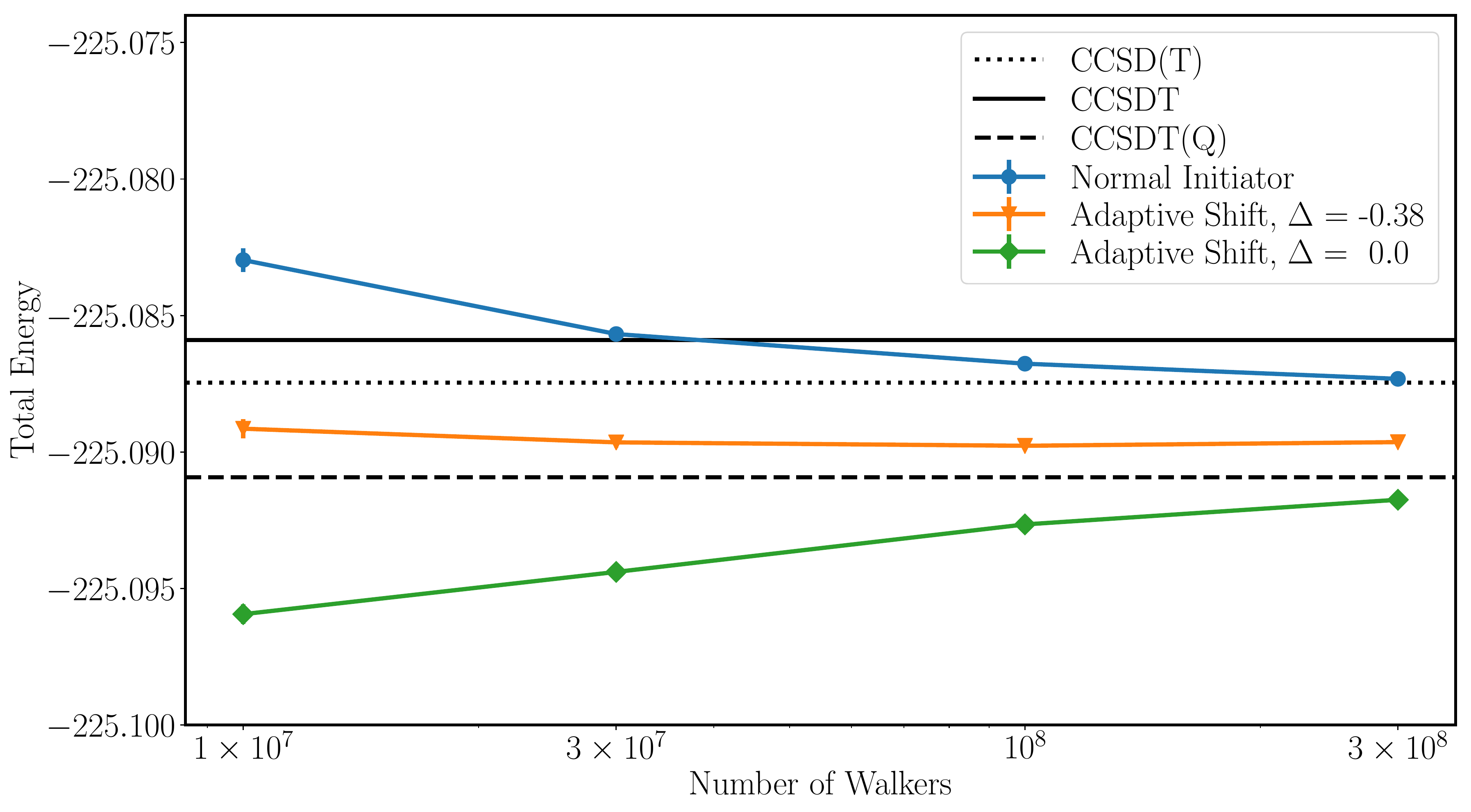}
		\caption{\label{fig:ozone_rm_vtz}
			Ozone in cc-pVTZ basis and RM geometry.
		}
		%\end{figure}
		%\begin{figure}
		\center
		\includegraphics[width=\columnwidth]{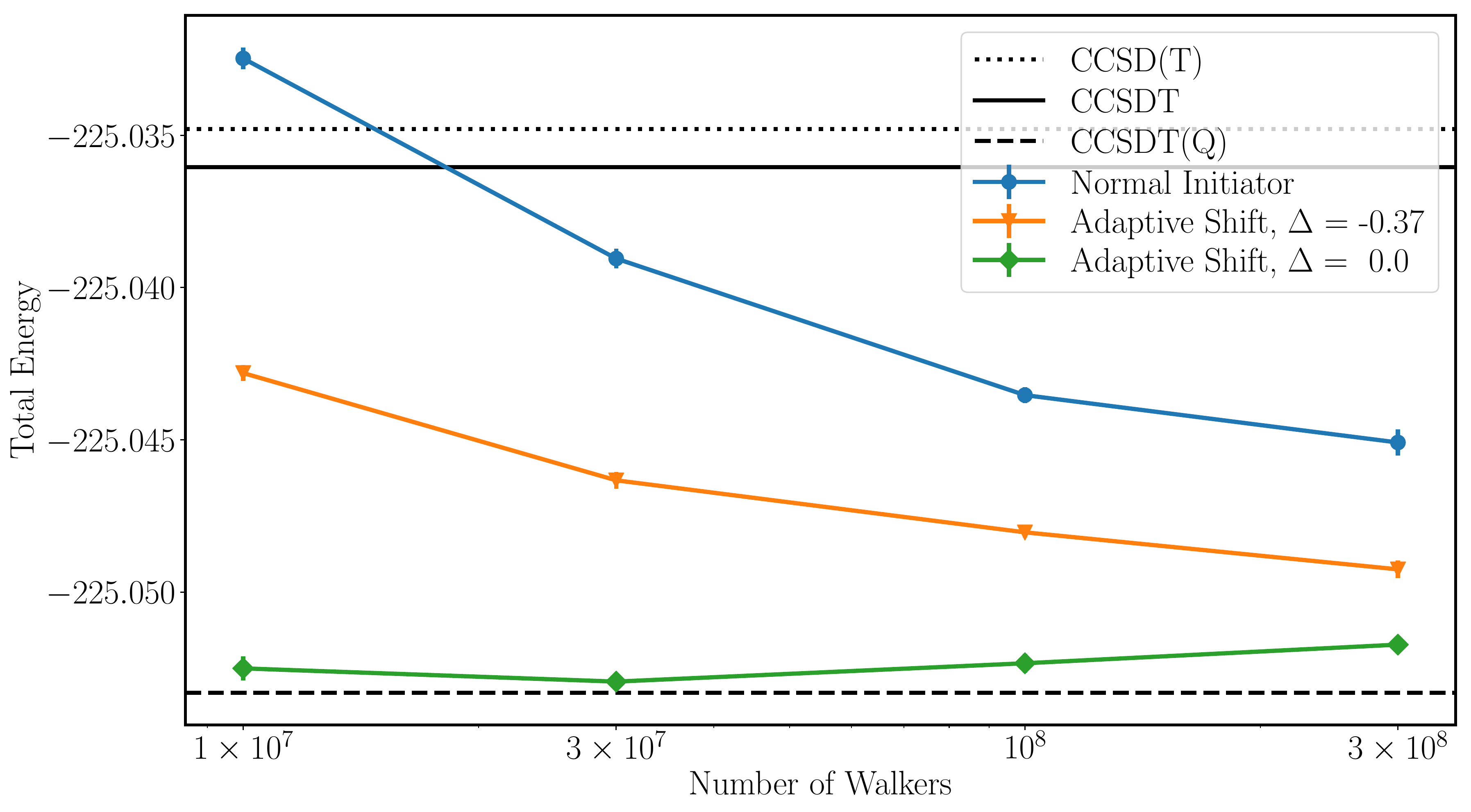}
		\caption{\label{fig:ozone_ts_vtz}
			Ozone in cc-pVTZ basis and TS geometry.
		}
	\end{figure}

	For the cc-pVTZ basis, we used populations of  $10$M, $30$M, $100$M, and $300$M walkers.
	For all geometries, the offset adaptive shift energy converges from above, with all values falling between CCSDT and CCSDT(Q) results.
	The converged energy of OM geometry is $-225.1375(1)$, which is  $5.8$ mH below the CCSDT result of $-225.1317$ and is $2.1$ mH above the CCSDT(Q) result of $-225.1396$ (Fig.~\ref{fig:ozone_om_vtz}).
	The converged energy of RM geometry is  $-225.0896(1)$ which is about $3.7$ mH below the CCSDT result of $-225.0859$  and about $1.3$ mH above the CCSDT(Q) result of $-225.0909$ (Fig.~\ref{fig:ozone_rm_vtz}).
	For the TS geometry, the full adaptive shift result increases up to $-225.0517(3)$, while the offset adaptive shift decrease down to $-225.0493(3)$. We believe the true value to be in between, about $1$ mH from our best estimate of $-225.050$.
	This value is  $3$ mH above the CCSDT(Q) result of $-225.0533$ and $14$ mH below the CCSDT result of $-225.0360$ (Fig.~\ref{fig:ozone_ts_vtz}).
	
\begin{figure}
		\center
		\includegraphics[width=\columnwidth]{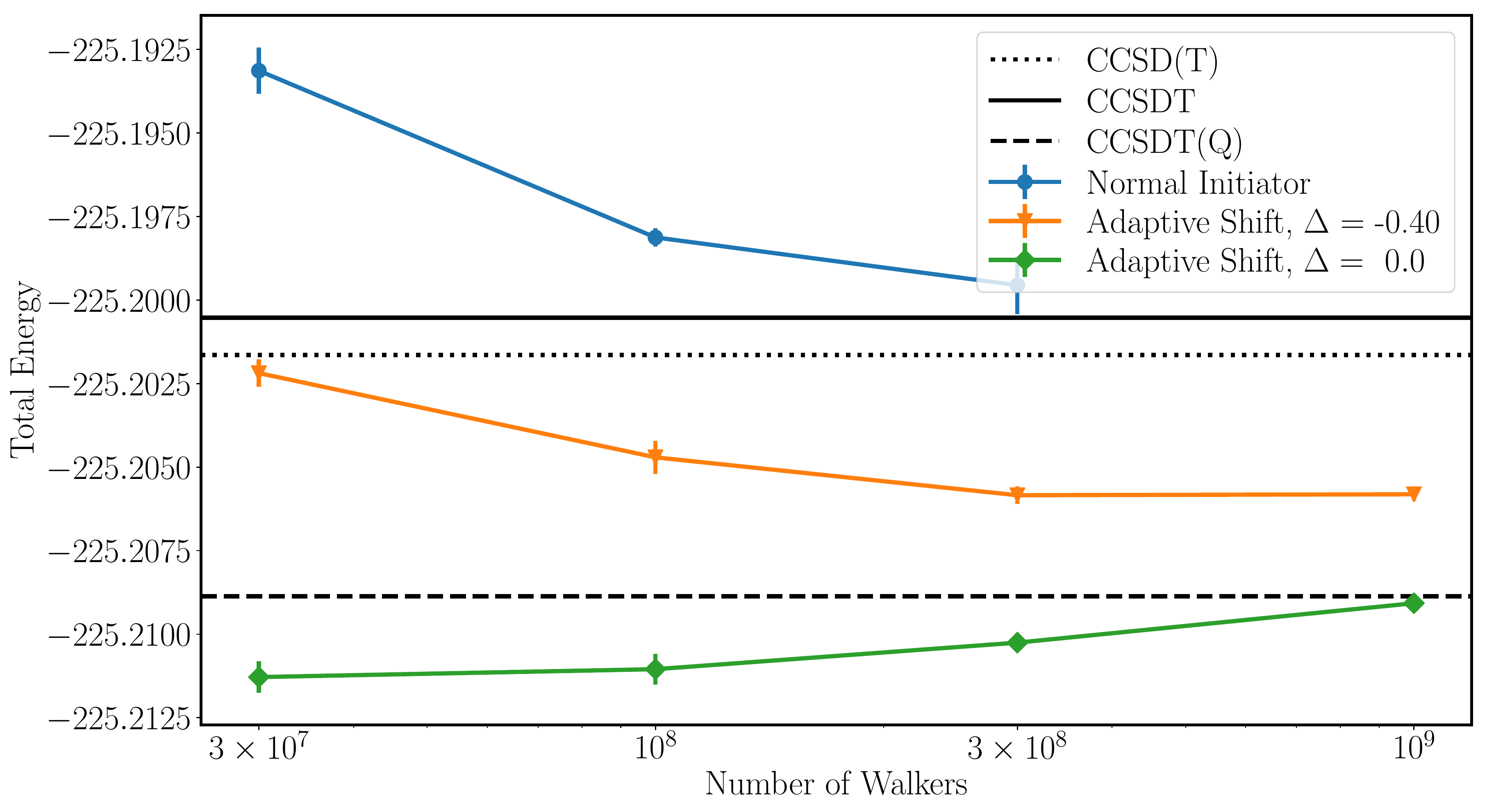}
		\caption{\label{fig:ozone_om_vqz}
			Ozone in cc-pVQZ basis and OM geometry.
		}
		%\end{figure}
		%\begin{figure}
		\center
		\includegraphics[width=\columnwidth]{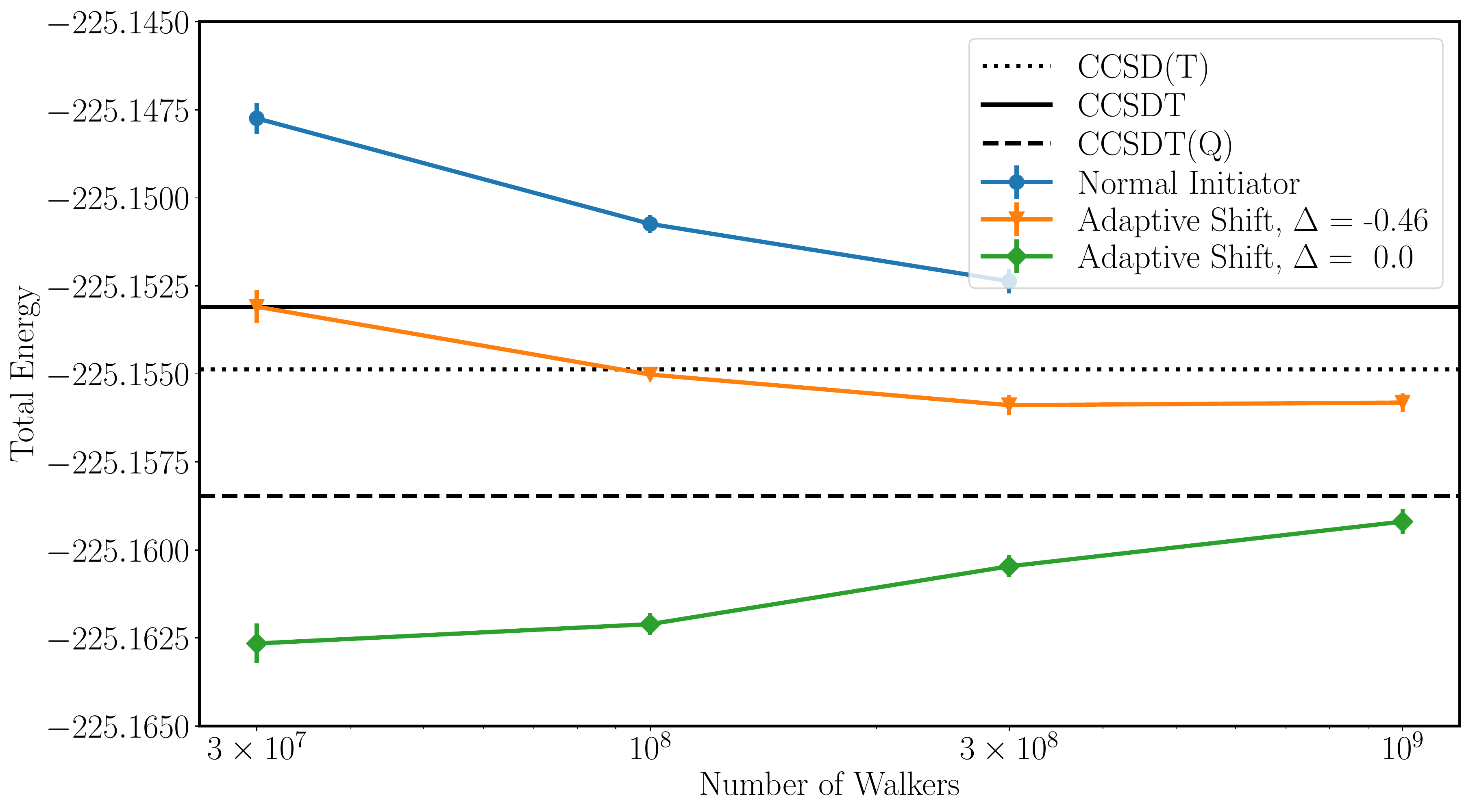}
		\caption{\label{fig:ozone_rm_vqz}
			Ozone in cc-pVQZ basis and RM geometry.
		}
		%\end{figure}
		%\begin{figure}
		\center
		\includegraphics[width=\columnwidth]{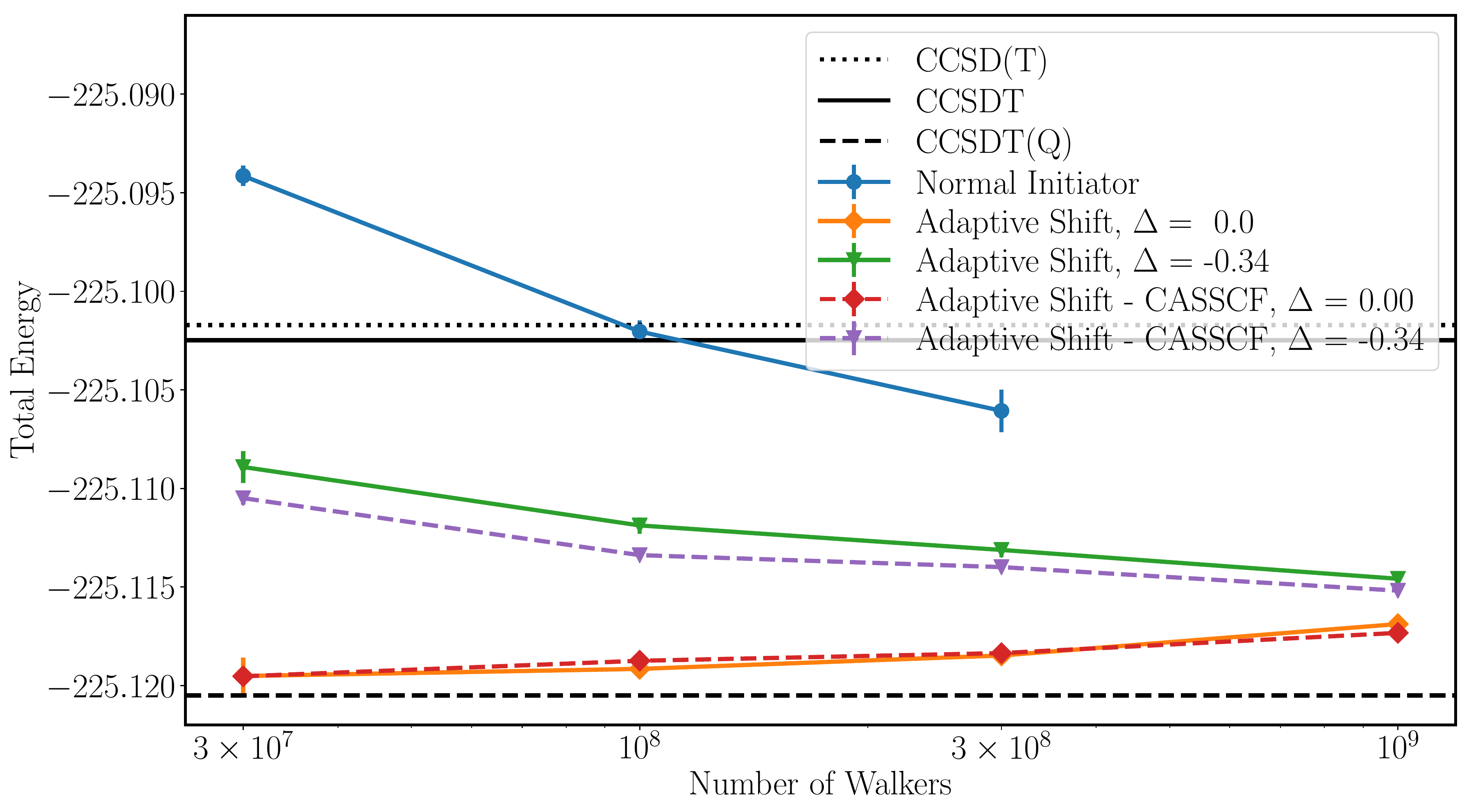}
		\caption{\label{fig:ozone_ts_vqz}
			Ozone in cc-pVQZ basis and TS geometry.
		}
	\end{figure}

	For the cc-pVQZ basis, we used  $30$M, $100$M, $300$M and $1$B walkers. 
	The offset adaptive shift energy for the OM geometry converges to $-225.2058(2)$, which is right between CCSDT result of $-225.2005$ and the CCSDT(Q) result of $-225.2089$ (Fig.~\ref{fig:ozone_om_vqz}).
	For the RM geometry, the offset adaptive shift energy converges from above to $-225.1558(3)$, which is $2.7$ mH below the CCSDT result of $-225.1531$ (Fig.~\ref{fig:ozone_rm_vqz}).
	For the TS geometry, the offset adaptive shift energy also converges from above down to the value $-225.1146(1)$, while the full adaptive shift converges from below up to $-225.1169(2)$. The midpoint $-225.116$ is about $13$ mH below the CCSDT result of $-225.1025$ (Fig.~\ref{fig:ozone_ts_vqz}).
	
	Interestingly, using CASSCF for the TS geometry in cc-pVQZ basis, we obtained molecular orbitals that give reference energy of $-224.2602$, $60$ mH below that of RHF energy of $-224.1998$.
	Using this new set of orbitals, we then applied the full adaptive shift and the offset adaptive shift with the same offset value as before.
	The full adaptive shift gives almost identical results to the ones using RHF orbitals with differences less than the statistical error bars.
	Using the offset the adaptive shift, the results are very close to the earlier ones and slightly below. The difference is about $1$ mH for $300$M walkers and gets smaller as the number of walkers increases.
	This demonstrates that the adaptive shift results are to a large extent independent of such changes of orbitals.

	To our knowledge, these are the first FCI estimates in a cc-pVQZ basis set of the valence-only correlated ozone molecule 
	in these three geometries. The gaps between these states can be compared with those reported by Umrigar et al using the semi-stochastic Heat-Bath CI-PT2 method at the cc-pVTZ level. At the cc-pVTZ level, the agreement between the two is on the order of less than 0.01eV, i.e. very good. The observed change in the gaps with basis set is much larger, highlighting the crucial importance of assessing basis-sets errors, to obtain quantitatively reliable energy differences. Since the memory footprint for treating large basis-sets can be prohibitive, the usefulness of the AS methodology is here shown, whereby the cc-pVQZ problem was treated with the same hardware resource as that for its cc-pVTZ counterpart. The Coupled Cluster gaps show erratic behavior with increasing the level of CC theory (with especially large changes observed in going from CCSDT to CCSDT(Q)), but an excellent agreement is achieved between the AS-FCIQMC gaps and CCSDTQ(P) gaps. Unfortunately, the latter was affordable only at the cc-pVDZ level, because of the memory requirements of storing the CCSDTQ method for larger basis sets.         
	
{ % begin box to localize effect of arraystretch change
\renewcommand{\arraystretch}{1.5}
	\begin{table*}[t]
	\begin{tabular}{@{\extracolsep{6pt}}l|l|l|l|l}
		Basis & Method       & OM-TS \  & OM-RM   & RM-TS\   \\ \hline \hline
		\multirow{5}{*} {cc-pVDZ}& CCSD(T) &$2.5537$ & $1.3761$& $1.1776$\\
		& CCSDT &$2.4872$ & $1.3931$ & $1.0941$\\
		& CCSDT(Q) &$2.2662$ & $1.4652$ & $0.801$\\ 
		& CCSDTQ &$2.3391$ & $1.4392$ & $0.8999$\\ 
		& CCSDTQ(P) &$2.2941$ & $1.4488$ & $0.8453$\\ 
		\cline{2-5}
		& AS-FCIQMC     & $2.313(1)$ &  $1.447(1)$ & $0.866(1)$\\ 
		
		\hline\hline
		\multirow{4}{*} {cc-pVTZ}& CCSD(T) & $2.6591$& $1.2262$ & $1.4329$ \\
		& CCSDT & $2.6029 $& $1.2463$ & $1.3566$ \\ 
		& CCSDT(Q) & $2.3472 $& $1.3235$ & $1.0237$\\ 	
		\cline{2-5}
		& SHCI & $2.41$ &  $1.30$ & $1.11$ \\ 
		\cline{2-5}
		& AS-FCIQMC& $2.40(3)$ & $1.302(4)$ & $1.10(3)$\\ 
		\hline\hline
		\multirow{3}{*} {cc-pVQZ}& CCSD(T) & $2.7193$ & $1.2727$ & $1.4466$ \\ 
		& CCSDT & $2.6681$ & $1.2909$ & $1.3772$\\ 
		& CCSDT(Q) & $2.4045$& $1.3714$ & $1.0331$\\ 	
		\cline{2-5}
		& AS-FCIQMC & $2.44(3)$ & $1.361(9)$ & $1.08(3)$ \\ \hline
	\end{tabular}
	\caption{\label{table:ozone_gaps}
		Energy gaps (in eV) between the different geometries of ozone. AS-FCIQMC values are computed using the offset adaptive shift results of the largest reported numbers of walkers. For cc-pVQZ basis in the TS geometry, we used our best estimate of $-225.116(1)$ mH, the midpoint between the full adaptive shift and the offset adaptive shift results using 1B walkers.}
	\mbox{}\\
    \end{table*}
}

	\subsection{Chromium Dimer}

	Next, we applied the offset adaptive shift FCIQMC method to the chromium dimer treated in a cc-pVDZ-DK basis with the X2C relativistic Hamiltonian, correlating 28 electrons in 76 orbitals 
	(Ne cores frozen).  We used orbitals obtained from a singlet-spin CASSCF(12,12) calculation.
	
\begin{figure}
		\center
		\includegraphics[width=\columnwidth]{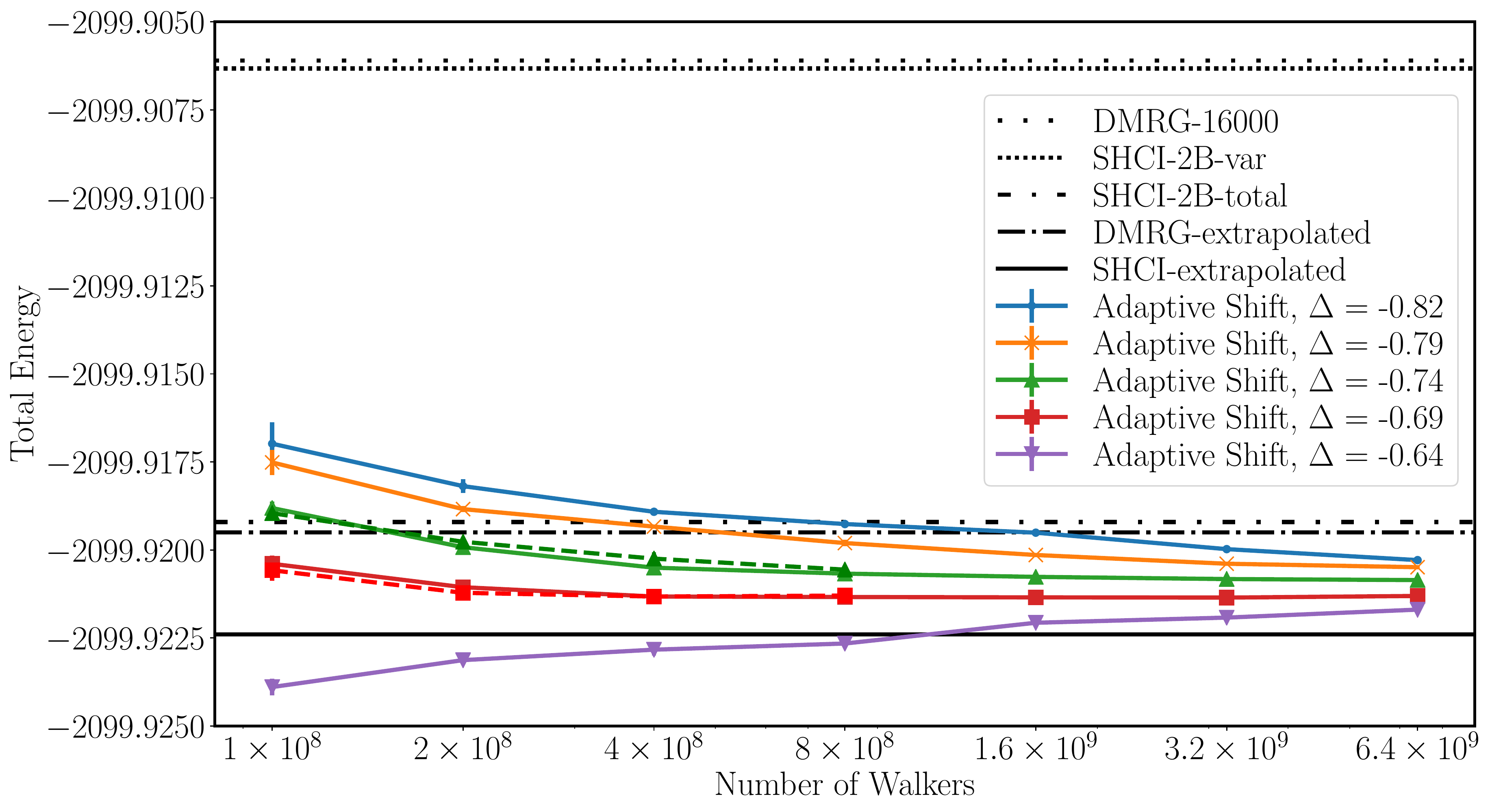}
		\caption{\label{fig:cr2}
			Chromium dimer in cc-pVDZ basis.
			Dashed lines represent adaptive shift results using the same offsets of the solid lines of the same color but with different seeds for the random number generator.
		}
	\end{figure}

	This system is a challenging multi-configurational problem even at the equilibrium geometry (1.68 \AA), and has been the subject of a recent extensive DMRG study going up to very large bond dimensions ~\cite{Guo2018}, as well as an SHCI study with very large variational spaces of 2-billion Slater determinants ~\cite{Li2018}. A DMRG variational total energy of  $-2099.9061$ is obtained using a bond dimension $M=16000$ in the default scheme, with an extrapolated FCI energy of $-2099.9195(27)$. The variational energy of SHCI is found to be $-2099.906322$. Adding the perturbative correction, the energy of the SHCI calculation falls to $-2099.919205$, and further extrapolation to the zero PT2 limit yields the FCI estimate of $-2099.9224(6)$ (see Fig.~\ref{fig:cr2}),  close to 3 mH below the DMRG estimate. 
	
	We started the calculations by using two offsets obtained as the correlation energies of two trial wavefunctions.
	The first offset $\Delta=-0.64$ is the correlation energy of the trial wavefunction resulting from diagonalizing the space of the top $100$ most-populated determinants in a normal initiator calculation.
	The other offset  $\Delta=-0.82$ is the correlation energy of the trial wavefunction obtained from diagonalizing the space of the top $1000$ most-populated determinants.
	In each case, the energies are estimated by projecting on the corresponding trial wavefunctions.
	We also used a semistochastic space with $|\mathcal{D}| = 50$K determinants.
	As we increase the number of walkers from $100$M up to $6.4$B, the energy using the first offset increases monotonically from $-2099.9239(2)$  to  $-2099.92170(5)$. 
	On the other hand, the energy using the second offset decreases monotonically from $-2099.9170(6)$ a.u. to $-2099.92029(3)$  bracketing the exact energy to within $1.4$ mH. 
	The results using $1.6$ billion walkers or larger fall between the extrapolated DMRG energy~\cite{Guo2018} of $-2099.9195(27)$ and the extrapolated SHCI energy~\cite{Li2018} of $-2099.9224(6)$ (see Fig.~\ref{fig:cr2}).

	To bracket the exact energy even more accurately, we ran additional adaptive shift calculations using three offsets evenly spaced between the previous two.
	At all walker populations except the largest ($6.4$B), we used a recent efficient implementation of the semistochastic method that allowed constructing very large semistochastic spaces with $|\mathcal{D}| = 1$M determinants.
	For $6.4$B walkers, we had to restrict the semistochastic space back to $50$K to avoid memory issues for these demanding calculations.
	The size of the trial space used in these calculations was set to $|\mathcal{T}| = 1000$.
	Using $\Delta=-0.74$, we get an energy estimate of  $-2099.92086(3)$ at $6.4$B walkers, while using $\Delta=-0.69$ we get a slightly lower estimate of  $-2099.92130(2)$  at the same number of walkers.
	Both these offsets appear to saturate for populations larger than $800$M walkers, so it is hard to tell which of the two provides a better estimate, although the discrepancy between them is small (0.4mH). 
	Combined with the trends of the other offsets, the true energy is likely to be between these two values, which pins it to within $0.3$ mH of our best estimate $-2099.921$.
	
	To make sure that these offset adaptive shift calculations are ergodic, we repeated the first four calculations using the same two offsets but with different seeds for the random number generator. The results of different seeds agree within the statistical error bars as shown in Fig.~\ref{fig:cr2}. 
    It is encouraging that, for the same offset, different simulation trajectories yield essentially identical energies, whilst the spread of energies among the different offsets is also not very large: about 0.5 mH - much smaller than the current discrepancy of 3 mH among the best alternative estimates given by DMRG or SHCI. Furthermore, the number of walkers required to give reliable energies is on the order of a few hundred million walkers - rather smaller than the 2B determinants used in the SHCI calculation. 

	\section{Summary and Discussion}
	In this paper, we have reviewed and extended the adaptive shift method for unbiasing the initiator approximation of FCIQMC.
	The adaptive shift method greatly helps to accelerate the convergence of the initiator approximation with respect to the number of walkers.	
	For some systems, the original adaptive shift method may overcorrect the initiator bias, leading to an overshoot in the energy estimates and a convergence from below.
	To remedy this, we have presented a simple way of modifying the adaptive shift correction that takes advantage of the inherent flexibility in the definition of the shift.
	We have shown that adding a fixed offset to the shift reduces the strength of the correction and provides a natural interpolation between the normal initiator method and the full adaptive shift method.
	In all cases where an offset is needed, we found that using a value, which is about half the correlation energy provides good results.
	The question of whether an optimal offset exists and how to automatically estimate it is still currently under investigation.	
	We observe a strict ordering between the energy curves of different offsets in all cases: lower offset leads to higher energy at any specific number of walkers.
	By varying the offset, we can achieve faster convergence to the exact energy and bracket it between the curves of different offsets.
	Using this approach, we are able to bin down energies of difficult strongly-correlated systems, such as chromium dimer in a cc-pVDZ basis, to a sub-mH accuracy.
    
    \section*{Supplementary Material}
	Tables of the results can be found in the supplementary material file.

	\section*{Acknowledgments}
	The authors gratefully acknowledge funding from the Max Planck Society. We thank Daniel Kats and Eugenio Vitale for useful discussions. 
	
	\section*{Data Availability}
	The data that supports the findings of this study are available within the article and its supplementary material.
	
	\section*{References}

\end{document}

\title{Supplementary Material - The Adaptive Shift Method in Full Configuration Interaction Quantum Monte Carlo: Development and Applications}

\author{Khaldoon Ghanem}
\affiliation{%
	Max Planck Institute for Solid State Research, Heisenbergstr. 1, 70569 Stuttgart, Germany
}
\email{k.ghanem@fkf.mpg.de}

\author{Kai Guther}
\affiliation{%
	Max Planck Institute for Solid State Research, Heisenbergstr. 1, 70569 Stuttgart, Germany
}

\author{Ali Alavi}
\affiliation{%
	Max Planck Institute for Solid State Research, Heisenbergstr. 1, 70569 Stuttgart, Germany
}
\affiliation{%
	Dept of Chemistry, University of Cambridge, Lensfield Road, Cambridge CB2 1EW, United Kingdom
}%

\date{\today}

\maketitle
	\begin{table*}[b]
	\begin{tabular}{@{\extracolsep{6pt}}r|l|l|l}
		$N_w/10^5$             & $3$                            & $10$                         & $30$                          \\ \hline \hline
		Normal i-FCIQMC      & $-109.278208(6)$ & $-109.278291(3)$ & $-109.278330(2)$ \\ \hline
		AS-FCIQMC, $\Delta=-0.25$ & $-109.278260(9)$ & $-109.278314(3)$ & $-109.278334(2)$ \\ \hline
		AS-FCIQMC, $\Delta=-0.20$ & $-109.278303(6)$ & $-109.278326(4)$ & $-109.278335(3)$ \\ \hline
		AS-FCIQMC, $\Delta=-0.15$ & $-109.278344(8)$ & $-109.278343(3)$ & $-109.278341(2)$ \\ \hline
		AS-FCIQMC, $\Delta=\ \ \;\; 0.0$ & $-109.278454(5)$ & $-109.278372(4)$ & $-109.278352(2)$ \\ \hline
		Exact	                       & \multicolumn{3}{r}{$-109.278339\quad\,$} \\ \hline
		HF                               & \multicolumn{3}{r}{$-108.949378\quad\,$} \\		
	\end{tabular}
	\caption{\label{table:n2_2.118_vdz}
		Total energy of N\textsubscript{2} in cc-pVDZ basis and near-equilibrium geometry: $\SI{2.118}{\bohr}$}
	\mbox{}\\
	%\end{table}
	%\begin{table}[]
	\begin{tabular}{@{\extracolsep{6pt}}r|l|l|l|l}
		$N_w/10^5$             & $3$                         & $10$                        & $30$                       & $100$                         \\ \hline \hline
		Normal i-FCIQMC      & $-108.96650(1)$ & $-108.96682(2)$ & $-108.966917(5)$ & $-108.966950(3)$ \\ \hline
		AS-FCIQMC, $\Delta=-0.55$ & $-108.96670(2)$ & $-108.96687(2)$ & $-108.966937(5)$ & $-108.966949(2)$ \\ \hline
		AS-FCIQMC, $\Delta=-0.43$ & $-108.96687(2)$ & $-108.96692(1)$ & $-108.966947(5)$ & $-108.966953(5)$ \\ \hline
		AS-FCIQMC, $\Delta=-0.28$ & $-108.96703(2)$ & $-108.96698(2)$ & $-108.966966(4)$ & $-108.966959(4)$ \\ \hline
		AS-FCIQMC, $\Delta=\ \ \;\; 0.0$ & $-108.96738(3)$ & $-108.96710(2)$ & $-108.966993(7)$ & $-108.966960(3)$ \\ \hline
		Exact	                       & \multicolumn{4}{r}{$-108.966950\quad\,$} \\ \hline
		HF                               & \multicolumn{4}{r}{$-108.222899\quad\,$} \\		
	\end{tabular}
	\caption{\label{table:n2_4.2_vdz}
		Total energy of N\textsubscript{2} in cc-pVDZ basis and stretched geometry: $\SI{4.2}{\bohr} $.}
	\mbox{}\\
	%\end{table}
	%\begin{table}[]
	\begin{tabular}{@{\extracolsep{6pt}}r|l|l|l|l}
		$N_w/10^6$             & $3$                         & $10$                        & $30$                       & $100$                       \\ \hline \hline
		Normal i-FCIQMC      & $-109.40522(2)$ & $-109.40544(1)$ & $-109.40560(1)$ & $-109.405721(6)$ \\ \hline
		AS-FCIQMC, $\Delta=-0.31$ & $-109.40550(3)$ & $-109.40560(1)$ & $-109.40566(1)$ & $-109.405761(7)$ \\ \hline
		AS-FCIQMC, $\Delta=-0.21$ & $-109.40577(4)$ & $-109.40573(2)$ & $-109.40575(2)$ & $-109.405799(7)$ \\ \hline
		AS-FCIQMC, $\Delta=-0.11$ & $-109.40604(2)$ & $-109.40587(2)$ & $-109.40585(1)$ & $-109.405821(6)$ \\ \hline
		AS-FCIQMC, $\Delta=\ \ \;\; 0.0$ & $-109.40637(4)$ & $-109.40604(1)$ & $-109.405890(8)$ & $-109.405863(9)$ \\ \hline
		CCSDT(Q)	              & \multicolumn{4}{r}{$-109.405885\quad\,$} \\
		CCSDTQ	                  & \multicolumn{4}{r}{$-109.405633\quad\,$} \\ \hline
		HF                               & \multicolumn{4}{r}{$-108.991084\quad\,$} \\		
	\end{tabular}
	\caption{\label{table:n2_eq_vqz}
		Total energy of N\textsubscript{2} in cc-pVQZ basis and equilibrium geometry: $\SI{1.0977 }{\angstrom}$.}
\end{table*}

	\begin{table*}[t]
	\begin{tabular}{@{\extracolsep{6pt}}r|l|l|l}
		$N_w/10^6$             & $10$                      & $30$                       & $100$                        \\ \hline \hline
		Normal i-FCIQMC      & $-224.9144(1)$   & $-224.91492(7)$ & $-224.91549(4)$ \\ \hline
		AS-FCIQMC, $\Delta=-0.34$ & $-224.91621(9)$ & $-224.91606(7)$ & $-224.91614(3)$ \\ \hline
		AS-FCIQMC, $\Delta=\ \ \;\; 0.0$ & $-224.91845(9)$ & $-224.91730(5)$ & $-224.91678(5)$ \\ \hline
		CCSDT(Q)                    & \multicolumn{3}{r}{$-224.91719\quad\,$} \\ \hline	
		CCSDTQ                    & \multicolumn{3}{r}{$-224.91555\quad\,$} \\ \hline
		CCSDTQ(P)               & \multicolumn{3}{r}{$-224.91641\quad\,$} \\ \hline		
		HF                               & \multicolumn{3}{r}{$-224.25933\quad\,$} \\		
	\end{tabular}
	\caption{\label{table:ozone_om_vdz}
		Total energy of Ozone in cc-pVDZ basis and open minimum geometry.}
	\mbox{}\\
	%\end{table}
	%\begin{table}[]
	\begin{tabular}{@{\extracolsep{6pt}}r|l|l|l}
		$N_w/10^6$             & $10$                        & $30$                      & $100$                        \\ \hline \hline
		Normal i-FCIQMC      & $-224.86147(6)$ & $-224.86183(4)$ & $-224.86238(2)$ \\ \hline
		AS-FCIQMC, $\Delta=-0.31$ & $-224.86320(7)$ & $-224.86293(3)$ & $-224.86297(4)$ \\ \hline
		AS-FCIQMC, $\Delta=\ \ \;\; 0.0$ & $-224.86498(6)$ & $-224.86414(5)$ & $-224.86357(2)$ \\ \hline
		CCSDT(Q)                    & \multicolumn{3}{r}{$-224.86334\quad\,$} \\ \hline
		CCSDTQ                    & \multicolumn{3}{r}{$-224.86266\quad\,$} \\ \hline
		CCSDTQ(P)               & \multicolumn{3}{r}{$-224.86317\quad\,$} \\ \hline		
		HF                               & \multicolumn{3}{r}{$-224.24251\quad\,$} \\		
	\end{tabular}
	\caption{\label{table:ozone_rm_vdz}
		Total energy of Ozone in cc-pVDZ basis and ring minimum geometry.}
	\mbox{}\\
	%\end{table}
	%\begin{table}[]
	\begin{tabular}{@{\extracolsep{6pt}}r|l|l|l}
		$N_w/10^6$             & $10$                    & $30$                      & $100$                        \\ \hline \hline
		Normal i-FCIQMC      & $-224.8283(1)$ & $-224.8293(1)$   & $-224.83019(5)$ \\ \hline
		AS-FCIQMC, $\Delta=-0.36$ & $-224.8309(1)$ & $-224.8310(1)$   & $-224.83115(4)$ \\ \hline
		AS-FCIQMC, $\Delta=\ \ \;\; 0.0$ & $-224.8336(1)$ & $-224.83282(9)$ & $-224.83210(8)$ \\ \hline
		CCSDT(Q)                    & \multicolumn{3}{r}{$-224.83391\quad\,$} \\ \hline	
		CCSDTQ                    & \multicolumn{3}{r}{$-224.82959\quad\,$} \\ \hline
		CCSDTQ(P)               & \multicolumn{3}{r}{$-224.83211\quad\,$} \\ \hline		
		HF                               & \multicolumn{3}{r}{$-224.11204\quad\,$} \\		
	\end{tabular}
	\caption{\label{table:ozone_ts_vdz}
		Total energy of Ozone in cc-pVDZ basis and transition state geometry.}
\end{table*}

	\begin{table*}[t]
	\begin{tabular}{@{\extracolsep{6pt}}r|l|l|l|l}
		$N_w/10^6$             & $10$                          & $30$                        & $100$      & $300$                   \\ \hline \hline
		Normal i-FCIQMC      & $-225.1281(3)$ & $-225.1319(1)$ & $-225.1342(1)$ & $-225.1349(1)$  \\ \hline
		AS-FCIQMC, $\Delta=-0.40$ & $-225.1348(3)$ & $-225.1364(2)$ & $-225.1374(2)$ & $-225.1375(1)$ \\ \hline
		AS-FCIQMC, $\Delta=\ \ \;\; 0.0$ & $-225.1436(6)$ & $-225.1422(2)$ & $-225.1414(2)$ & $-225.1400(3)$ \\ \hline
		CCSD(T)                        & \multicolumn{4}{r}{$-225.1325\quad\,$} \\ \hline	
		CCSDT                        & \multicolumn{4}{r}{$-225.1317\quad\,$} \\ \hline
		CCSDT(Q)                  & \multicolumn{4}{r}{$-225.1396\quad\,$} \\ \hline		
		HF                               & \multicolumn{4}{r}{$-224.3339\quad\,$} \\		
	\end{tabular}
	\caption{\label{table:ozone_om_vtz}
		Total energy of Ozone in cc-pVTZ basis and open minimum geometry.}
	\mbox{}\\
	%\end{table}
	%\begin{table}[]
	\begin{tabular}{@{\extracolsep{6pt}}r|l|l|l|l}
		$N_w/10^6$             & $10$                          & $30$                        & $100$  & $300$                        \\ \hline \hline
		Normal i-FCIQMC      & $-225.0830(4)$ & $-225.0857(1)$ & $-225.0868(1)$ & $-225.0873(2)$  \\ \hline
		AS-FCIQMC, $\Delta=-0.38$ & $-225.0891(3)$ & $-225.0896(2)$ & $-225.0898(1)$ & $-225.0896(1)$ \\ \hline
		AS-FCIQMC, $\Delta=\ \ \;\; 0.0$ & $-225.0959(4)$ & $-225.0944(2)$ & $-225.0927(1)$ & $-225.0917(1)$ \\ \hline
		CCSD(T)                        & \multicolumn{4}{r}{$-225.0875\quad\,$} \\ \hline			
		CCSDT                        & \multicolumn{4}{r}{$-225.0859\quad\,$} \\ \hline
		CCSDT(Q)                  & \multicolumn{4}{r}{$-225.0909\quad\,$} \\ \hline		
		HF                               & \multicolumn{4}{r}{$-224.3188\quad\,$} \\		
	\end{tabular}
	\caption{\label{table:ozone_rm_vtz}
		Total energy of Ozone in cc-pVTZ basis and ring minimum geometry.}
	\mbox{}\\
	%\end{table}
	%\begin{table}[]
	\begin{tabular}{@{\extracolsep{6pt}}r|l|l|l|l}
		$N_w/10^6$             & $10$                          & $30$                        & $100$   & $300$     \\ \hline \hline
		Normal i-FCIQMC      & $-225.0325(4)$ & $-225.0390(3)$ & $-225.0435(3)$ & $-225.0451(4)$  \\ \hline
		AS-FCIQMC, $\Delta=-0.37$ & $-225.0428(3)$ & $-225.0463(3)$ & $-225.0480(2)$ & $-225.0493(3)$ \\ \hline
		AS-FCIQMC, $\Delta=\ \ \;\; 0.0$ & $-225.0525(4)$ & $-225.0529(3)$ & $-225.0523(3)$ & $-225.0517(3)$ \\ \hline
		CCSD(T)                        & \multicolumn{4}{r}{$-225.0348\quad\,$} \\ \hline			
		CCSDT                        & \multicolumn{4}{r}{$-225.0360\quad\,$} \\ \hline
		CCSDT(Q)                  & \multicolumn{4}{r}{$-225.0533\quad\,$} \\ \hline		
		HF                               & \multicolumn{4}{r}{$-224.1829\quad\,$} \\		
	\end{tabular}
	\caption{\label{table:ozone_ts_vtz}
		Total energy of Ozone in cc-pVTZ basis and transition state geometry.}
\end{table*}

	\begin{table*}[t]
	\begin{tabular}{@{\extracolsep{6pt}}r|l|l|l|l}
		$N_w/10^6$ & $30$ & $100$ & $300$   & $1000$ \\ \hline \hline
		Normal i-FCIQMC      & $-225.1931(7)$ & $-225.1981(3)$ & $-225.1996(9)$ &  \\ \hline
		AS-FCIQMC, $\Delta=-0.40$ & $-225.2022(4)$ & $-225.2047(5)$ & $-225.2058(3)$ & $-225.2058(2)$\\ \hline
		AS-FCIQMC, $\Delta=\ \ \;\; 0.0$ & $-225.2113(5)$ & $-225.2110(5)$ & $-225.2103(2)$ & $-225.2091(1)$  \\ \hline
		CCSD(T)                        & \multicolumn{4}{r}{$-225.2016\quad\,$} \\ \hline		
		CCSDT                        & \multicolumn{4}{r}{$-225.2005\quad\,$} \\ \hline
		CCSDT(Q)                  & \multicolumn{4}{r}{$-225.2089\quad\,$} \\ \hline		
		HF                               & \multicolumn{4}{r}{$-224.3518\quad\,$} \\		
	\end{tabular}
	\caption{\label{table:ozone_om_vqz}
		Total energy of Ozone in cc-pVQZ basis and open minimum geometry.}
	\mbox{}\\
	%\end{table}
	%\begin{table}[]
	\begin{tabular}{@{\extracolsep{6pt}}r|l|l|l|l}
		$N_w/10^6$ & $30$ & $100$ & $300$   & $1000$ \\ \hline \hline
		Normal i-FCIQMC      & $-225.1477(4)$ & $-225.1507(3)$ & $-225.1524(3)$ & \\ \hline
		AS-FCIQMC, $\Delta=-0.46$ & $-225.1531(5)$ & $-225.1550(1)$ & $-225.1559(3)$ & $-225.1558(3)$\\ \hline
		AS-FCIQMC, $\Delta=\ \ \;\; 0.0$ & $-225.1627(6)$ & $-225.1621(3)$ & $-225.1605(3)$ & $-225.1592(4)$ \\ \hline
		CCSD(T)                      & \multicolumn{4}{r}{$-225.1549\quad\,$} \\ \hline
		CCSDT                        & \multicolumn{4}{r}{$-225.1531\quad\,$} \\ \hline
		CCSDT(Q)                  & \multicolumn{4}{r}{$-225.1585\quad\,$} \\ \hline		
		HF                               & \multicolumn{4}{r}{$-224.3351\quad\,$} \\		
	\end{tabular}
	\caption{\label{table:ozone_rm_vqz}
		Total energy of Ozone in cc-pVQZ basis and ring minimum geometry.}
	\mbox{}\\
	%\end{table}
	%\begin{table}[]
	\begin{tabular}{@{\extracolsep{6pt}}r|l|l|l|l}
		$N_w/10^6$ & $30$ & $100$ & $300$   & $1000$ \\ \hline \hline
		Normal i-FCIQMC      & $-225.0941(5)$ & $-225.1020(6)$ & $-225.106(1)$ &  \\ \hline
		AS-FCIQMC, $\Delta=-0.34$ & $-225.1089(8)$ & $-225.1119(4)$ & $-225.1131(4)$ & $-225.1146(1)$ \\ \hline
		AS-FCIQMC (CASSCF),  $\Delta=-0.34$ & $-225.1105(3)$ & $-225.1134(2)$ & $-225.1140(3)$ & $-225.1152(2)$ \\ \hline
		AS-FCIQMC, $\Delta=\ \ \;\; 0.0$ & $-225.1195(9)$ & $-225.1192(2)$ & $-225.1185(4)$ & $-225.1169(2)$ \\ \hline
		AS-FCIQMC (CASSCF), $\Delta=\ \ \;\; 0.0$ & $-225.1195(4)$ & $-225.1187(3)$ & $-225.1184(2)$ & $-225.1173(3)$\\ \hline
		CCSD(T)                     & \multicolumn{4}{r}{$-225.1017\quad\,$} \\ \hline
		CCSDT                        & \multicolumn{4}{r}{$-225.1025\quad\,$} \\ \hline
		CCSDT(Q)                  & \multicolumn{4}{r}{$-225.1205\quad\,$} \\ \hline		
		HF                               & \multicolumn{4}{r}{$-224.1998\quad\,$} \\		
	\end{tabular}
	\caption{\label{table:ozone_ts_vqz}
		Total energy of Ozone in cc-pVQZ basis and transition state geometry.}
\end{table*}

	\begin{turnpage}
	\begin{table*}[p]
		\begin{tabular}{@{\extracolsep{6pt}}r|l|l|l|l|l|l|l}
			$N_w/10^8$             & $1$                          & $2$                        & $4$      & $8$    & $16$  & $32$  & $64$   \\ \hline \hline
			Normal i-FCIQMC      & $-2099.90380(5)$ & & & & & & \\ \hline
			AS-FCIQMC, $\Delta=-0.82$ & $-2099.9170(6)$ & $-2099.9182(2)$ & $-2099.9189(1)$ & $-2099.9193(1)$ & $-2099.91951(5)$ & $-2099.91998(5)$ & $-2099.92029(3)$ \\ \hline
			AS-FCIQMC, $\Delta=-0.79$ & $-2099.9175(4)$ & $-2099.91884(9)$ & $-2099.91934(5)$ & $-2099.91980(9)$ & $-2099.92014(4)$ & $-2099.92039(2)$ & $-2099.92049(3)$\\ \hline		
			AS-FCIQMC, $\Delta=-0.74$ & $-2099.9188(2)$ & $-2099.91992(8)$ & $-2099.92050(5)$ & $-2099.92067(8)$ & $-2099.92076(3)$ & $-2099.92083(2)$ & $-2099.92086(3)$ \\ \hline		
			AS-FCIQMC, $\Delta=-0.69$ & $-2099.9204(2)$ & $-2099.92106(7)$ & $-2099.9213(1)$ & $-2099.92134(4)$ & $-2099.92135(4)$ & $-2099.92135(3)$ & $-2099.92130(2)$\\ \hline								
			AS-FCIQMC, $\Delta=-0.64$ & $-2099.9239(2)$ & $-2099.92313(9)$ & $-2099.9228(2)$ & $-2099.92266(7)$ & $-2099.92207(7)$ & $-2099.92192(5)$ & $-2099.92170(5)$ \\ \hline
			AS-FCIQMC, $\Delta=\ \ \, 0.00$ & $-2099.9480(4)$ & & & & & & \\ \hline
			DMRG-16000                               & \multicolumn{7}{r}{$-2099.9061\ \,\quad\,$} \\	\hline
			SHCI-2B-var                               & \multicolumn{7}{r}{$-2099.906322\ \ $} \\	\hline	
			SHCI-2B-total                               & \multicolumn{7}{r}{$-2099.919205\ \ $} \\	\hline
			DMRG-extrapolated                              & \multicolumn{7}{r}{$-2099.9195(27)$} \\	\hline			
			SHCI-extrapolated                               & \multicolumn{7}{r}{$-2099.9224(6)\ \,$} \\	\hline				
			HF                               & \multicolumn{7}{r}{$-2098.51188\quad\,$} \\		
		\end{tabular}
		\caption{\label{table:cr2}
			Total energy of chromium dimer in cc-pVDZ basis.}
		\mbox{}\\
	\end{table*}
\end{turnpage}